\documentclass[review,5p,fleqn]{elsarticle}

\usepackage[T1]{fontenc}
\usepackage{epsfig}
\usepackage{graphicx}
\usepackage{amsmath}
\usepackage{amsfonts}
\usepackage{amssymb}
\usepackage{floatflt}

\newtheorem{thm}{Theorem}

\newdefinition{rmk}{Remark}
\newproof{pf}{Proof}
\newproof{pot}{Proof of Theorem \ref{thm2}}

\journal{Journal of Computational and Theoretical Chemistry}
\begin{document}
\begin{frontmatter}
\title{Construction and counting of the number of operators  included in a normalized vibrational Hamiltonian with $n$
degrees of freedom with a $p:q$ resonance}
\author[Dijon]{G.Saget}
\author[Dijon,Tomsk]{C. Leroy \corref{cor}}
\cortext[cor]{Corresponding author.}
\ead{claude.leroy@u-bourgogne.fr}
\author[Dijon]{H.R. Jauslin}
\address[Dijon]{Laboratoire Interdisciplinaire Carnot de Bourgogne (ICB),
UMR 6303 CNRS-Universit\'{e} Bourgogne Franche-Comt\'{e} 9 Av. A.
Savary, BP 47 870, F-21078 DIJON Cedex, FRANCE.}
\address[Tomsk]{Tomsk Polytechnic University, Institute of Physics and Technology, Department of General Physics, 30 Lenin Avenue, 634050 Tomsk, RUSSIA.}

\begin{abstract}
\label{res} This paper is the second one of two papers aimed at
constructing hamiltonian systems of n degrees of freedom. In
molecular spectroscopy, the construction of vibrational Hamiltonians
for highly excited molecular systems through algebraic formalism
implies to introduce "by hand" the operators  reflecting the
exchanges of quanta of energy between the different oscillators. It
is thus tedious to predict, for any given order of the development
of the Hamiltonian, the total number of operators which should
appear in the Hamiltonian (\cite{JOY1997}, \cite{JJS1999},
\cite{JFS2002}). In this second paper, we propose a method of
construction of a normalized vibrational Hamiltonian of a highly
excited molecular system with  $n$ degrees of freedom in the case of
a a $p:q$ resonance. We present also the counting of all the
independent operators and the counting of all the parameters
included in the Hamiltonian (Counting theorems $1$ to $8$). The
method introduces, on a systematic way, all the operators, in
particular the coupling operators, that can be built from the
polynomials formed by products of powers of the generators of a Lie
algebra: the algebra of the invariant polynomials built in classical
mechanics from the the kernel $Ker \,ad_{\mathcal{H}_{0}}$ of the
adjoint operator $ad_{\mathcal{H}_{0}}$ (see \cite{SLJ2015} or
\cite{BHLV2003}, \cite{MHO2009}). Application to the non-linear
triatomic molecule ClOH is then given, taking into account the Fermi
resonance between the O-Cl stretching oscillators and the bending
motion. The study of this molecular system in highly excited
vibrational states (until almost the dissociation limit) has been
realized in \cite{JJS1999}, with  a fit of 725 levels of energy. On
the 86 coefficients (among which 31 coupling coefficients) that we
count, and completely compatible with \cite{JJS1999}, the smallest
rms value leads to keep only 28 non-zero coefficients. In the
appendix, we explain the vocabulary and the strategy employed in
order to demonstrate the theorems of coupling operators included in
the Hamiltonian.
\end{abstract}
\end{frontmatter}

\section{Introduction} \label{intro}
This second article is the continuation of a first one \cite{SLJ2015} where we applied an algebraic approach to study highly excited molecular systems
with no resonance between two of the oscillators representing the molecular system.

After some basic reminders about the normalization (\cite{BIR1927},
\cite{GUS1964}, \cite{DEP1969}), we built a vibrational Hamiltonian
written as a Dunham expansion on the basis of the generators of the
invariant polynomial algebra and we have counted all the operators
included in the Hamiltonian developed until the order $N$ (Eq.
$(13)$ of \cite{SLJ2015}). An application to the non-linear
triatomic molecule of ClOH has been performed as the highly excited
vibrational states of this molecule have been widely studied
(\cite{WAY1991}, \cite{SBP1998}, \cite{SBM1999}). Thus the
Hamiltonian  (Table $2$ of \cite{SLJ2015}) allows to reproduce the
vibrational structure of 314 energy levels (until  $70$ $\%$ of the
dissociation energy) with a set of $34$ coefficients which is in
good agreement with \cite{JJS1999}. In order to describe the
vibrational structure of more excited levels, it is necessary to
take into account a $2:1$ resonance between the stretching
oscillator associated with the  O-Cl bond and the bending oscillator
associated to the angular deformation between bonds O-Cl and O-H.
The coupling operators are then implemented by hand on a more or
less arbitrary manner and it is a laborious task to predict in
advance the total number of operators appearing in the Hamiltonian.

In the present paper, after some definitions and reminders about the standard normalization of the harmonic oscillator of quadratic part  $\mathcal{H}_{0}$
(section \ref{norm}), we present a building method of the normalized vibrational Hamiltonian for a molecular system having two degrees of freedom fulfilling
a  $p:q$ resonance condition ($p$ and $q$ are positif integers with a gcd equal to 1).
We derive also the enumeration of all the independent operators and the number of involved coefficients. Then we extend the method to systems having  $n$ degrees of freedom (section \ref{ham.norm}). The method introduces on a systematic way all the operators, in particular the coupling operators, which can be obtained from polynomial formed by products of powers of the generators of the invariant polynomial algebra. Then we establish the theorems counting the coupling
operators involved in the Hamiltonian
 (sections \ref{Th2.mon} and \ref{Th3.mon}).
Demonstration of these theorems imply to implement a specific
vocabulary which is given in Appendix. We end our paper with the
counting of all the operators and parameters involved in the
normalized Hamiltonian describing the vibrational structure of ClOH
molecule until $98$ $\%$ of the dissociation limit (section
\ref{Appli}).

\section{Normalization}\label{norm} \subsection{Lowest order
Hamiltonian: $\mathcal{H}_{0}$}\label{ham} Suppose
$\mathcal{H}(q_{1},\,...,\, q_{n},\, p_{1},\,..., \, p_{n})$ the
classical vibrational Hamiltonian of an Hamiltonian system with  $n$
degrees of freedom, the quadratic part of which is an anisotropic
harmonic oscillator $\mathcal{H}_{0} =
\sum_{k=1}^{n}\frac{\omega_{k}(q_{k}^{2}+p_{k}^{2})}{2}$, where the
$n$ quantities  $\omega_{k}$ are characteristic  pulsations of the
oscillators (with $1 \leq i,\,j \leq n$, $\omega_{i} \neq
\omega_{j}$) and $q_{k}$, $p_{k}$ are the canonical dimensionless
variables generalized coordinates and conjugate momenta defined on
the phase space $\Gamma$.

\subsection{Definition of the $p:q$ resonance}\label{reson} The
hamiltonian $\mathcal{H}_{0}$ shows a unique $p:q$ resonance if two
pulsations  (refer here-after as $\omega_{1}$ and $\omega_{2}$) are
connected through a relation of the form:
\begin{eqnarray}\label{eq.1} \frac{\omega_{2}}{\omega_{1}} =
\frac{p}{q}, \end{eqnarray} $p$ and $q$ are two positif integers
with gcd($p$, $q$) = 1 and $p\geq q$ \cite{BHLV2003}, the $n-2$
others pulsations $\omega_{k}$ satisfying to the condition of non
resonance (Equation $(3)$ of \cite{SLJ2015}).

\subsection{Equations of motion}\label{Equations} We introduce the
complex variables  $z_{k}$ and $z_{k}^{\ast}$ defined in function of
the canonical variables $q_{k}$ and $p_{k}$ ($1\leq k \leq n$) as:
$z_{k}  = \frac{1}{\sqrt{2}}(q_{k}+ip_{k})$ and $z_{k}^{\ast}  =
\frac{1}{\sqrt{2}}(q_{k}-ip_{k})$.  The lowest order Hamiltonian can
be now rewritten  $\mathcal{H}_{0}=
-\imath(\omega_{1}z_{1}{z_{1}}^{\ast}+\frac{p}{q}\omega_{1}
z_{2}{z_{2}}^{\ast}+\sum_{k=3}^{n}\omega_{k}z_{k}{z_{k}}^{\ast})$.
With these new variables $z_{k}$, the equations of motion reads:
\begin{equation}\label{eq.2} \frac{d z_{1}}{dt}   = -\imath
\omega_{1} z_{1},\, \frac{d z_{2}}{dt}  = -\imath
\frac{p}{q}\omega_{1} z_{2},\,\frac{d z_{k}}{dt} = -\imath
\omega_{k} z_{k}\,\,(3 \leq k \leq n). \end{equation}

\subsection{Hamiltonian flow }\label{Flow} From an initial condition
$z_{0}=(z_{1,0},\,...,\,z_{n,0})$, formally the solution of the
equations of motion is written as $z(t)=
{\phi_{t}}^{\mathcal{H}_{0}}(z_{0})$.
${\phi_{t}}^{\mathcal{H}_{0}}:\Gamma \rightarrow \Gamma$ is the
Hamiltonian flow generated by $\mathcal{H}_{0}$ (\cite{ARN1976,
CB1997, EFS2005})
 \\
We have: $z(t) = {\phi_{t}}^{\mathcal{H}_{0}}(z_{0}) \Longrightarrow
\left(
                                                    \begin{array}{c}
                                                      z_{1}(t) \\
                                                      \vdots \\
                                                      z_{n}(t) \\
                                                    \end{array}
                                                  \right)$
\begin{eqnarray}\label{eq.3}
=\left(
   \begin{array}{cccc}
     e^{-\imath \omega_{1}t} & 0 & \ldots & 0  \\
     0 & e^{-\imath \frac{p}{q}\omega_{1}t} &  \ldots & 0 \\
     0 & 0 & \ddots & 0 \\
     0 & 0 & 0 & e^{-\imath \omega_{n}t}
   \end{array}
 \right)
 \left(
                                              \begin{array}{c}
                                                z_{1,0} \\
                                                \vdots \\
                                                z_{n,0} \\
                                              \end{array}
                                            \right).
\end{eqnarray}
Solutions $z(t)$ are the trajectories of the Hamiltonian flow or the orbits of the harmonic oscillator  with a $p:q$ resonance.

\subsection{Hilbert basis}\label{Base}
The condition of resonance  (\ref{eq.1}) implies that the kernel $Ker
\,ad_{\mathcal{H}_{0}}$ of the adjoint operator
$ad_{\mathcal{H}_{0}}$, (defined by  equations $(4)$ and $(5)$ of
\cite{SLJ2015}), is generated  by the $n+2$ monomials
(\cite{BHLV2003}) ($n \geq 2$ integer) :
\begin{equation}\label{eq.4}
\sigma_{-1} = {{{z}_{1}}^{\ast}}^{p}{{z}_{2}}^{q}, \sigma_{0} =
{{z}_{1}^{p}}{{{z}_{2}}^{\ast}}^{q}, \sigma_{1} =
{{z}_{1}}{{z}_{1}}^{\ast},\,...,\, \sigma_{n} =
{{z}_{n}}{{z}_{n}}^{\ast}.
\end{equation}

The kernel has the structure of a Lie algebra, called algebra of the
invariant polynomials. The generators $\sigma_{k}$ ($-1\leq k \leq
n$) form a basis of  $Ker$ $ad_{\mathcal{H}_{0}}$ called Hilbert
basis (\cite{HIL1893, GAT2000, CLO1992}). As in the non resonant
case, the generators of the Hilbert basis are invariant under the
flow of the harmonic oscillator  $\phi_{t}^{\mathcal{H}_{0}}$, which
is a symplectic symmetry for the generators \cite{SLJ2015}. While
for the case $n=2$, \cite{EFS2005} used
 $J=p\sigma_{1}+q\sigma_{2}$,
$\Pi_{1}=p\sigma_{1}-q\sigma_{2}$,
$\Pi_{2}=\sqrt{2^{p+q}p^{q}q^{p}}Re\biggl({{{z}_{1}}^{\ast}}^{p}{{z}_{2}}^{q}\biggr)$
and
$\Pi_{3}=\sqrt{2^{p+q}p^{q}q^{p}}Im\biggl({{{z}_{1}}^{\ast}}^{p}{{z}_{2}}^{q}\biggr)$,
we prefer to conserve equations (\ref{eq.4}) which are more easy to use in order to build a normalized quantum Hamiltonian
as we will see in section \ref{ham.norm}.

\subsection{Poisson brackets of the generators}\label{croch} The
Poisson brackets of the generators are equal to zero except:  $\{
\sigma_{-1}, \,\sigma_{1} \}= ip\sigma_{-1}$, $\{ \sigma_{-1},
\,\sigma_{2} \}= -iq\sigma_{-1}$, $\{ \sigma_{0}, \,\sigma_{1} \}=
-ip\sigma_{0}$, $\{ \sigma_{0}, \,\sigma_{2} \}= iq\sigma_{0}$ and
\\ $\{ \sigma_{-1}, \,\sigma_{0} \}=
i{\sigma_{1}}^{p-1}{\sigma_{2}}^{q-1}(p^{2}\sigma_{2}-q^{2}\sigma_{1})$.

\subsection{Reduced phase space}\label{esp} The generators of the
invariant polynomials algebra, $\sigma_{-1}$, $\sigma_{0}$,
$\sigma_{1}$ and $\sigma_{2}$ ($\sigma_{1} \geq 0$, $\sigma_{2} \geq
0$), are not independent. They satisfy the relation:
\begin{eqnarray}\label{eq.5}
(\frac{\sigma_{0}+\sigma_{-1}}{2})^{2}+(\frac{\sigma_{0}-\sigma_{-1}}{2\imath})^{2}
={\sigma_{1}}^{p}{\sigma_{2}}^{q}. \end{eqnarray} In the phase space
$\Gamma = \mathbb{R}^{2n}$, the iso-$\mathcal{K}$-energy surfaces
are hyper-surfaces of $\mathbb{R}^{2n-1}$.\\
 For a given value $h_{0}$ of
$\mathcal{H}'_{0}=-\frac{\mathcal{H}_{0}}{\imath}$ (obviously $h_{0}
> 0$), in the case of a  $p:q$ resonance, (\ref{eq.5}) becomes:

\begin{eqnarray}\label{eq.6}
{\sigma'_{0}}^{2}+{\sigma'_{-1}}^{2}
={\sigma_{1}}^{p}\biggl(\frac{h_{0}}{\omega_{2}}-\frac{q}{p}\sigma_{1}-\sum_{k=3}^{n}(\frac{\omega_{k}}{\omega_{2}})\sigma_{k}\biggr)^{q},
\end{eqnarray}
with $\sigma'_{0}=\frac{\sigma_{0}+\sigma_{-1}}{2}$ and
$\sigma'_{-1}=\frac{\sigma_{0}-\sigma_{-1}}{2\imath}$ and the condition $h_{0} \geq
\omega_{2}\frac{q}{p}\sigma_{2}+\sum_{k=3}^{n}\omega_{k}\sigma_{k}$.
In the phase space, Eq. (\ref{eq.6}) defines the reduced phase space \cite{EFS2005}, on which the dynamics of the motion
is reduced to a space of dimension $n$.

\begin{figure}[h]
\begin{center}
\epsfig{file=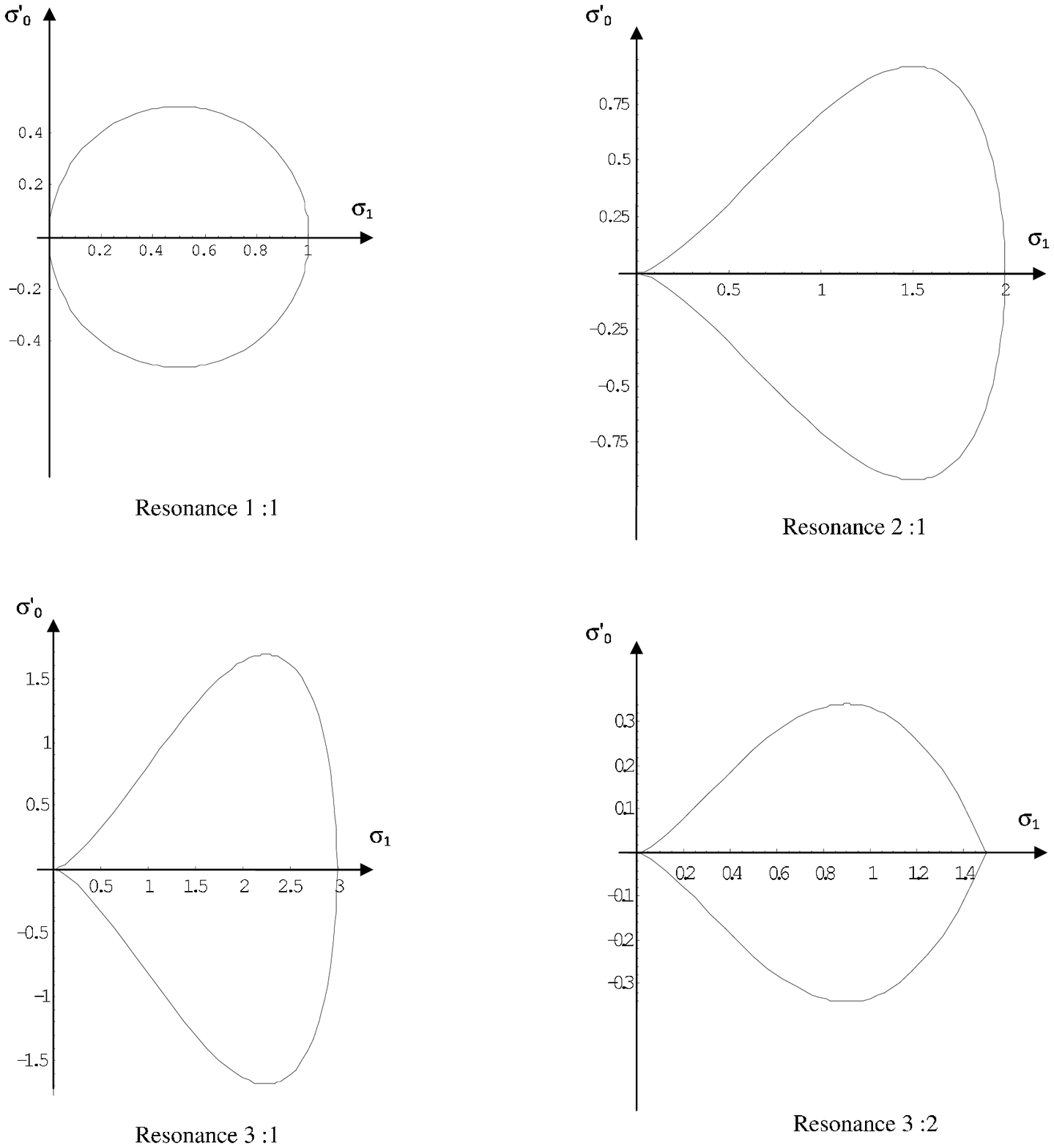,height=11cm, width=9cm}
\end{center}
\caption{In the plan  $(\sigma_{1}$, $\sigma'_{0})$ for
$\frac{h_{0}}{\omega_{2}} = 1$ S.I.: reduced phase space for different
$p:q$ resonance values.}
\end{figure}

\subsection{The normalized Hamiltonian $\mathcal{K}$} We want to
define a normalized Hamiltonian $\mathcal{K}$ verifying
$\{\mathcal{H}_{0},\,\mathcal{K}\} =
-ad_{\mathcal{H}_{0}}(\mathcal{K})=ad_{\mathcal{K}}(\mathcal{H}_{0})=0$.
$\mathcal{K}$ can be written on the form (\cite{BHLV2003}):
$\mathcal{K} = \mathcal{H}_{0} +
f(\sigma_{-1},\,\sigma_{0},\,\sigma_{1},\,...,\, \sigma_{n})$,
moreover $\mathcal{K}$ has to be invariant under time reversal
operation (TRO). Generators $\tau(\sigma_{k})=\sigma_{k}$ for $1\leq
k \leq n$ are also invariant under TRO except $\sigma_{-1}$ and
$\sigma_{0}$ which verify $ \sigma_{-1} \leftarrow $ TRO
$\rightarrow \sigma_{0}$.

\section{Construction of the normalized Hamiltonian}\label{ham.norm}
\subsection{Case $n=2$}
\subsubsection{Development of the Hamiltonian on the Hilbert basis}
We show this case as an example of modeling of the vibrational
normalized Hamiltonian describing high excited stretching modes of
triatomic ABC or AB$_{2}$ molecules. The resonant normalized
Hamiltonian $\mathcal{K}$, with quadratic term $\mathcal{H}_{0} =
-\imath \frac{\omega_{1}}{q}(q\sigma_{1}+p\sigma_{2})$ is only
function of $\sigma_{-1}$, $\sigma_{0}$, $\sigma_{1}$ and
$\sigma_{2}$: $\mathcal{K} = \mathcal{H}_{0} +
f(\sigma_{-1},\,\sigma_{0},\,\sigma_{1},\, \sigma_{2})$. We expand
the normalized Hamiltonian as a polynomial development of powers of
the generators until an order $N$. In what follows, $N \geq p+q+4$,
$q_{0}$, $q_{1}$, $q_{2}$, $q_{3}$, $\delta$ and $\beta$ are
integers; values of $Q_{0}$, $Q_{1}$, $Q_{2}$ and $Q_{3}$ in the
sums have to be precise in order that the development based on the
generators of the algebra be effectively a polynomial expansion.
>From \cite{SLJ2015} we already know that $Q_{0}=E(\frac{N}{2})$. In
the expression of $\mathcal{K}$ here-after, we will need to
demonstrate that $Q_{1}=E(\frac{N}{p+q})$, $Q_{2}= E(\frac{\delta
-(p+q)}{2})$ and $Q_{3}= E(\frac{\beta -(p+q)}{2})$:
\begin{eqnarray}\label{eq.7}
\mathcal{K} & = & \mathcal{H}_{0} + \sum_{q_{0} =
2}^{Q_{0}}({\alpha_{q_{0}}^{1}} {\sigma_{1}}^{q_{0}} +
{\alpha_{q_{0}}^{2}} {\sigma_{2}}^{q_{0}})  \notag \\
&+& \sum_{r=2}^{Q_{0}}\sum_{1\leq i_{1} < i_{2} \leq
2}\sum_{r_{i_{1}} \geq 1,\,r_{i_{2}}\geq1}^{r_{i_{1}}+r_{i_{2}} = r}
{\alpha_{r_{i_{1}},\,r_{i_{2}}}^{i_{1},\,i_{2}}}
{\sigma_{i_{1}}^{r_{i_{1}}}}{\sigma_{i_{2}}^{r_{i_{2}}}} \notag
\\
&+& \sum_{q_{1}=1}^{Q_{1}}
({\alpha_{q_{1}}^{0}}{\sigma_{-1}^{q_{1}}}+{{\alpha'}_{q_{1}}^{0}}{\sigma_{0}^{q_{1}}}) \notag \\
& + & \sum_{\delta=p+q+2}^{N}\sum_{q_{2}=1}^{Q_{2}}
\biggl({\alpha_{E(\frac{\delta-2q_{2}}{p+q}),\,q_{2}}^{-1,\,1}}{\sigma_{-1}^{E(\frac{\delta-2q_{2}}{p+q})}}{\sigma_{1}^{q_{2}}} \notag \\
& + &
{{\alpha}_{E(\frac{\delta-2q_{2}}{p+q}),\,q_{2}}^{0,\,1}}{\sigma_{0}^{E(\frac{\delta-2q_{2}}{p+q})}}{\sigma_{1}^{q_{2}}}\biggr) \notag \\
& + & \sum_{\delta=p+q+2}^{N}\sum_{q_{2}=1}^{Q_{2}}
\biggl({\alpha_{E(\frac{\delta-2q_{2}}{p+q}),\,q_{2}}^{-1,\,2}}{\sigma_{-1}^{E(\frac{\delta-2q_{2}}{p+q})}}{\sigma_{2}^{q_{2}}} \notag \\
& + &
{{\alpha'}_{E(\frac{\delta-2q_{2}}{p+q}),\,q_{2}}^{0,\,2}}{\sigma_{0}^{E(\frac{\delta-2q_{2}}{p+q})}}{\sigma_{2}^{q_{2}}}\biggr) \notag \\
& + &
\sum_{\beta=p+q+4}^{N}\sum_{q_{3}=2}^{Q_{3}}\sum_{\gamma=1}^{q_{3}-1}
\biggl({\alpha_{E(\frac{\beta-2q_{3}}{p+q}),\,\gamma,\,q_{3}-\gamma}^{-1,\,1,\,2}}{\sigma_{-1}^{E(\frac{\beta-2q_{3}}{p+q})}}
{\sigma_{1}^{\gamma}{\sigma_{2}^{q_{3}-\gamma}}} \notag \\
& + &
{{\alpha}_{E(\frac{\beta-2q_{3}}{p+q}),\,\gamma\,q_{3}-\gamma}^{0,\,1,\,2}}{\sigma_{0}^{E(\frac{\beta-2q_{3}}{p+q})}}
{\sigma_{1}^{\gamma}{\sigma_{2}^{q_{3}-\gamma}}}\biggr).
\end{eqnarray}

In section \ref{ind}, we will prove that the invariance of
$\mathcal{K}$ under TRO implies that all the coefficients of
$\mathcal{K}$ are purely  imaginary.\\ Furthermore, one
distinguishes in $\mathcal{K}$ the monomials built as powers of
$\sigma_{-1}$ or $\sigma_{0}$ and we will call them coupling
monomials between the two resonant oscillators : they appear in the
sums $S_{0}=\sum_{q_{1}=1}^{Q_{1}}
({\alpha_{q_{1}}^{0}}{\sigma_{-1}^{q_{1}}}+{{\alpha'}_{q_{1}}^{0}}{\sigma_{0}^{q_{1}}})$,
\\ $S_{m,\,\ell}^{(2)}=\sum_{\delta=p+q+2}^{N}\sum_{q_{2}=1}^{Q_{2}}
{\alpha_{E(\frac{\delta-2q_{2}}{p+q}),\,q_{2}}^{m,\,\ell}}{\sigma_{m}^{E(\frac{\delta-2q_{2}}{p+q})}}{\sigma_{\ell}^{q_{2}}}$
or $S_{m}^{(3)}$ $=$
$\sum_{\beta=p+q+4}^{N}\sum_{q_{3}=2}^{Q_{3}}\sum_{\gamma=1}^{q_{3}-1}
{\alpha_{E(\frac{\beta-2q_{3}}{p+q}),\,\gamma,\,q_{3}-\gamma}^{m,\,1,\,2}}{\sigma_{m}^{E(\frac{\beta-2q_{3}}{p+q})}}
{\sigma_{1}^{\gamma}{\sigma_{2}^{q_{3}-\gamma}}}$. ($m=-1,\,0$;
$\ell=1,\,2$) \\ Others monomials are the non resonant monomials.

\subsubsection{Invariance of $\mathcal{K}$ under TRO}\label{inv}
Hypothesis of the invariance of the Hamiltonian $\mathcal{K}$ under TRO implies the following identification:
\begin{eqnarray}\label{eq.8}
{\alpha_{q_{1}}^{0}} &=& {{\alpha'}_{q_{1}}^{0}} \notag \\
{\alpha_{E(\frac{\delta-2q_{2}}{p+q}),\,q_{2}}^{-1,\,\ell}}
&=&{{\alpha}_{E(\frac{\delta-2q_{2}}{p+q}),\,q_{2}}^{0,\,\ell}}\,\, (\ell=1,\,2) \notag \\
{\alpha_{E(\frac{\beta-2q_{3}}{p+q}),\,\gamma,\,q_{3}-\gamma}^{-1,\,1,\,2}}&=&
{{\alpha}_{E(\frac{\beta-2q_{3}}{p+q}),\,\gamma,\,q_{3}-\gamma}^{0,\,1,\,2}}.
\end{eqnarray}
Taking into account the relations Eq. (\ref{eq.8}), one may factorize sets of  $\lambda$-monomials of same coefficients in Eq. (\ref{eq.7}).
These $\lambda$-monomials are all real, in particular the following coupling monomials:
$\biggl({\sigma_{-1}^{E(\frac{\delta-2q_{2}}{p+q})}}+{\sigma_{0}^{E(\frac{\delta-2q_{2}}{p+q})}}\biggr){\sigma_{\ell}^{q_{2}}}$
and $\biggl({\sigma_{0}^{E(\frac{\beta-2q_{3}}{p+q})}}
+{\sigma_{-1}^{E(\frac{\beta-2q_{3}}{p+q})}}\biggr)
{\sigma_{1}^{\gamma}{\sigma_{2}^{q_{3}-\gamma}}}$. Moreover the transformation of the $(q_{k},\,p_{k})$ in $(z_{k},\,z_{k}^{\ast})$ is symplectic of multiplier  $-i$, like the Hamiltonian $\mathcal{K}$. One thus deduce that all the coefficients in Eq. (\ref{eq.7}) are purely imaginary.

\subsubsection{Independence of the coefficients }\label{ind}
>From the Jacobi identity $(-1\leq j \leq 2)$:
$\{\mathcal{K},\,\{\sigma_{j},\,\mathcal{H}_{0}\}\}+\{\mathcal{H}_{0},$
$\,\{\mathcal{K},\,\sigma_{j}\}\}+
\{\sigma_{j},\,\{\mathcal{H}_{0},\,\mathcal{K}\}\}=0$ and calculating all the Poisson brackets  $\{\mathcal{K},\,\sigma_{j}\}$, knowing that
 $(\mathcal{K},\,\sigma_{j}) \in Ker$
$ad_{\mathcal{H}_{0}}$, one deduces that none relation exists between the coefficients of the development of Eq. (\ref{eq.7}):  the different $\lambda$-monomials are independent between them.

\subsubsection{Range of variation of the $Q_{i}$}\label{toto} We
will now precise the range of variation of the integers appearing in
(\ref{eq.7}). \begin{itemize}
\item For $Q_{1}$: the monomials
$\sigma_{-1}^{q_{1}}$ and $\sigma_{0}^{q_{1}}$ are of degree
$d'=(p+q)q_{1}$. $q_{1}$ takes all integer values from $1$ to the
maximal value $Q_{1}$ satisfying ($N \geq p+q$): $Q_{1} =
E(\frac{N}{p+q})$. If $N < p+q$, all the coefficients
${\alpha_{q_{1}}^{0}}$ are equal to zero.
\item For $Q_{2}$: the
$2$-monomials ${\sigma_{-1}^{p_{2}}}{\sigma_{1}^{q_{2}}}$,
${\sigma_{0}^{p_{2}}}{\sigma_{1}^{q_{2}}}$,
${\sigma_{-1}^{p_{2}}}{\sigma_{2}^{q_{2}}}$ and
${\sigma_{0}^{p_{2}}}{\sigma_{2}^{q_{2}}}$ ($p_{2}$ positive
integer) are of degree $(p+q)p_{2}+2q_{2}$. In an expansion of
$\mathcal{K}$ to a given order $\delta$, the following condition
should be verified for the degree of the different $2$-monomials:
\begin{eqnarray}\label{eq.9}
(p+q)p_{2}+2q_{2} = \delta.
\end{eqnarray}
As the products of the powers of these  $2$-monomials
should always appear in $\mathcal{K}$, it implies $p_{2} \geq 1$ and
$q_{2} \geq 1$, i.e. $\delta \geq p+q+2$. The integer $q_{2}$ may
vary from  1 to a maximal value $Q_{2}$, obtained for the smallest
value of $p_{2}$ ($p_{2} = 1$). From (\ref{eq.9}), one deduces that
$Q_{2} = E(\frac{\delta-(p+q)}{2})$. \\ We determine now the range
of variation of the integer $p_{2}$. Conditions $(\theequation)$ and
$p_{2}$ integer imply that, for a fixed value of $\delta$ , with
$1\leq q_{2} \leq Q_{2}$, $p_{2}$ varies between the integer values
of $E(\frac{\delta-2Q_{2}}{p+q})$ and $E(\frac{\delta-2}{p+q})$. In
particular, for $\delta=p+q+2$ or $p+q+3$: $Q_{2}=1$ and $p_{2}=1$;
however for $\delta=p+q+3$, the condition (\ref{eq.9}) is not
satisfied. This will not be detrimental to the counting of the
coefficients in the sums $S_{m,\,\ell}^{(2)}$ but will introduce
redundancies among the $2$-monomials (See paragraph \ref{multi})
which will then be eliminated (See appendix). Thereafter, we denote
$p_{2}=E(\frac{\delta-2q_{2}}{p+q})$. \\ Finally $\delta$ varies
from $p+q+2$ to $N$ ($N\geq p+q+2$). If $\delta < p+q +2$, all the
coefficients in the sums involving $2$-monomials are equal to zero.
\item For $Q_{3}$: the $3$-monomials
${\sigma_{-1}^{p_{3}}}{\sigma_{1}^{\gamma}}{\sigma_{2}^{r_{3}}}$ and
${\sigma_{0}^{p_{3}}}{\sigma_{1}^{\gamma}}{\sigma_{2}^{r_{3}}}$
($p_{3}$, $\gamma$ and $r_{3}$ are positive integers) are of degree
$(p+q)p_{3}+2\gamma+2r_{3}$. As for the $2$-monomials, one requires
that the following condition is respected:
\begin{eqnarray}\label{eq.10} (p+q)p_{3}+2\gamma+2r_{3}=\beta.
\end{eqnarray}
As the products of powers of the $3$-monomials have to be include in
$\mathcal{K}$: $p_{3} \geq 1$, $\gamma \geq 1$ and $r_{3} \geq 1$,
thus $\beta \geq p+q+4$. We denote $q_{3}=\gamma+r_{3}$. For a fixed
value of $q_{3}$, $\gamma$ takes all the values from 1 to $q_{3}-1$.
Furthermore we impose no constraint to the integer $q_{3}$ which
takes all values from  2 ($r_{3}=\gamma =1$) to a maximal value
$Q_{3}$, obtained for $p_{3} = 1$. With (\ref{eq.10}), we find that
$Q_{3} = E(\frac{\beta-(p+q)}{2})$. \\ We determine now the range of
variation of the integer  $p_{3}$. For a fixed value of $\beta$,
conditions $(\theequation)$ and $p_{3}$ integer imply that, for
$2\leq q_{3} \leq Q_{3}$, $p_{3}$ varies between the integer values
of $E(\frac{\beta-2Q_{3}}{p+q})$ and $E(\frac{\beta-4}{p+q})$. In
particular, for $\beta=p+q+4$ or $\beta=p+q+5$: $Q_{3}=2$ and
$p_{3}=1$.  For $\beta=p+q+5$,  condition (\ref{eq.10}) is not
fulfilled, thus the counting of the coefficients in the sums
$S_{m}^{(3)}$ will show redundancies among the $3$-monomials (See
section \ref{multi}) that will have to be removed (See appendix).
Thereafter, we denote $p_{3}=E(\frac{\beta-2q_{3}}{p+q})$. All the
coefficients in the sums involving $3$-monomials are \textit{a
priori} non equal to zero except if $p+q+4 \leq \beta \leq N$.
\end{itemize}

\subsubsection{Counting}\label{denomb} For a given value of $N$,
$2E(\frac{N}{p+q})$ monomials appear in the sum $S_{0}$. For given
values of $N$ and $\delta$, a sum $S_{m,\,\ell}^{(2)}$ contains
$Q_{2}$ $2$-monomials. Thus one deduces that it contains
$\Lambda_{1} = \sum_{\delta = p+q+2}^{N}E(\frac{\delta -(p+q)}{2})$
$2$-monomials. Denoting $N = p+q+2+K$ with $K$ integer and $\delta =
p+q+2+k$ with $k$ an integer such that $k = 0,\,...,\,K$:
$\Lambda_{1} = \sum_{k=0}^{K}E(1+\frac{k}{2})$. The sum contains
$K+1$ monomials but the calculation of  $\Lambda_{1}$ depends on the
parity of $K$: \begin{itemize} \item if $K=2\tilde{p}$ ($\tilde{p}$
integer) is even, thus for $k = 0,\,...,\,K$, each integer in the
sum appears two times: 1 associated to $E(1)$, $E(3/2),\,...$,
except the last term giving $E(\tilde{p}+1) = \tilde{p}+1$ which
appears only one time, thus we get: \begin{eqnarray}\label{eq.11}
\Lambda_{1} = \frac{1}{4}[N-(p+q)]^{2}. \end{eqnarray} \item if
$K=2\tilde{p}+1$ ($\tilde{p}$ integer) is odd, thus for $k =
0,\,...,\,K$, each integer in the sum appears two times: 1
associated to $E(1)$, $E(3/2),\,...,\,\tilde{p}+1$ for
$E(\tilde{p}+1)$ and $E(\tilde{p}+3/2)$. After calculations, one
obtains: \begin{eqnarray}\label{eq.12} \Lambda_{1} =
\frac{1}{4}[N-(p+q)-1][N-(p+q)+1]. \end{eqnarray} \end{itemize} For
fixed values of $N$ and $\beta$, a sum $S_{m}^{(3)}$ contains
$\sum_{q_{3}=2}^{Q_{3}}(q_{3}-1)=\frac{Q_{3}(Q_{3}-1)}{2}$ . This
sum has \\ $\Lambda_{2} = \sum_{\beta =
p+q+4}^{N}\frac{\biggl(E(\frac{\beta-(p+q)}{2})\biggr)\biggl(E(\frac{\beta-(p+q)}{2})-1\biggr)}{2}$
$3$-monomials. We denote now $N = p+q+4+K$ with $K$ integer and
$\beta = p+q+4+k$ with $k$ integer such that $k = 0,\,...,\,K$,
$\Lambda_{2} =
\sum_{k=0}^{K}\frac{\biggl(E(2+\frac{k}{2})\biggr)\biggl(E(2+\frac{k}{2})-1\biggr)}{2}$.
Also we denote $\tilde{S}_{1}= \sum_{k=0}^{K}E^{2}(2+\frac{k}{2})$
and $\tilde{S}_{2}= \sum_{k=0}^{K}E(2+\frac{k}{2})$. The calculation
of these two sums depends on the parity of $K$: \begin{itemize}
\item if $K=2\tilde{p}$ ($\tilde{p}$ integer) is even, then for $k =
0,\,...,\,K$, the following integers appear two times in the sum
$\tilde{S_{1}}$: 4 associated to $E^{2}(2)$, $E^{2}(5/2)$, 9 for
$E^{2}(3)$, $E^{2}(7/2),\,...$; only the last term giving
$E^{2}(\tilde{p}+2) = (\tilde{p}+2)^{2}$ appears once. With the same
process applied for $\tilde{S}_{2}$, one finally obtains:
$\tilde{S}_{1}=\frac{(\tilde{p}+1)(2\tilde{p}^{2}+4\tilde{p}+3)}{3}$
and $\tilde{S}_{2}= (\tilde{p}+1)(\tilde{p}+2)+\tilde{p}$.
\end{itemize} Thus one deduces that: \begin{eqnarray}\label{eq.13}
\Lambda_{2} = \frac{1}{24}[N-(p+q)-2][N-(p+q)-1][N-(p+q)].
\end{eqnarray} \begin{itemize} \item if $K=2\tilde{p}+1$
($\tilde{p}$ integer) is odd, for $k = 0,\,...,\,K$, the following
integers appear two times in the sum $\tilde{S_{1}}$: 4 associated
to $E^{2}(2)$, $E^{2}(5/2)$, 9 for $E^{2}(3)$,
$E^{2}(7/2),\,...,\,(\tilde{p}+2)^{2}$ for $E^{2}(\tilde{p}+2)$ and
$E^{2}(\tilde{p}+5/2)$. Performing all the calculations, it reads:
\\
$\tilde{S}_{1}=\frac{(\tilde{p}+2)(\tilde{p}+3)(2\tilde{p}+5)}{3}-2$
and $\tilde{S}_{2}= (\tilde{p}+2)(\tilde{p}+3)-2$. \end{itemize}
Finally, the result is given by: \begin{eqnarray}\label{eq.14}
\Lambda_{2} = \frac{1}{24}[N-(p+q)-3][N-(p+q)-1][N-(p+q)+1].
\end{eqnarray} The number of others monomials involved in the
Hamiltonian $\mathcal{K}$ is given by
$\Lambda=\sum_{k=1}^{2}C_{2}^{k}C_{Q_{0}}^{k}=\frac{Q_{0}(Q_{0}+3)}{2}$
(Eq. $(13)$ of \cite{SLJ2015} for  $n=2$).

\subsection{Dissatisfaction of the counting} Eq. (\ref{eq.11}) to
(\ref{eq.14}) take into account the redundancy of some $2$-monomials
or $3$-monomials in the expression of (\ref{eq.7}). For instance,
for a $1:1$ resonance, developing $\mathcal{K}$ until $N=7$ ($K=3$),
Eq. (\ref{eq.12}) predicts  six $2$-monomials in a sum
$S_{m,\,\ell}^{(2)}$ instead of three in reality.

\subsection{Multiplicity of some $\lambda$-monomials}\label{multi}
\subsubsection{Pointing the problem} In the formula (\ref{eq.7}), we
associate the couple of integers $C=(p_{2},\,q_{2})$ ($2$-couple) or
$(p_{3},\,q_{3},\,\gamma)$ ($3$-couple) to a $2$-monomial or a
$3$-monomial.  In fact, in the next section, we will count the
$\lambda$-monomials in the sums $S_{m,\,\ell}^{(2)}$  and
$S_{m}^{(3)}$ ($m=-1,\,0$; $\ell=1,\,2$) step by step from $\delta =
p+q+2$ or $\beta=p+q+4$ until the desired value of $N$ ($N \geq
p+q+2$) by eliminating  the $2$-monomials and $3$-monomials which
are redundant. Firstly, we will define the necessary definitions and
tools in the sections \ref{def} until \ref{calcul}.

\subsubsection{Definitions}\label{def} \begin{itemize}
\item We call
main interval, any set of values taken by the integer $N$, order of
development of $\mathcal{K}$, between two integers $N_{1}$ and
$N_{2}$ ($N_{2}> N_{1}$): $N\in [N_{1},\,N_{2}]$. The smallest
possible value of $N$ in the relation (\ref{eq.7}) is $N=p+q+2$ for
the $2$-monomials and $p+q+4$ for the $3$-monomials. \item For fixed
$\delta$ (respectively $\beta$), we call secondary interval  $IS$,
the set of all the values taken by the integer $q_{2}$ (respectively
$q_{3}$) between the integers $1$ (respectively $2$) and
$Q_{2}=E(\frac{\delta -(p+q)}{2})$ (respectively
$Q_{3}=E(\frac{\beta -(p+q)}{2})$). We denote $IS_{\delta}=
[1,\,Q_{2}]_{\delta}$ or $IS_{\beta}=[2,\,Q_{3}]_{\beta}$ or even
more simply $IS$.
\item For fixed $\delta$ or $\beta$, we will say
that a couple of integers $C$ is present if it appears in the
secondary interval $IS$. In the opposite case, it will be declared
absent.
\item A couple of integers  $C$ is said present on a main
interval $[N_{1},\,N_{2}]$ if it appears at least one time on one of
the secondary intervals $N_{2}-N_{1}+1$, each of these secondary
intervals being constituted from one of the $N_{2}-N_{1}+1$ values
of  $\delta$ or $\beta$ composing the main interval. If $C$ does not
appear on all the secondary intervals, it will be said as absent on
the main interval $[N_{1},\,N_{2}]$.
\item We define the
multiplicity $\mu$ of a couple  $C$ as being the number of times
this couple appears in an given interval (main or secondary). \item
The cumulative multiplicity (also denoted by $\mu$) of a couple $C$
on a secondary interval $IS$ in the number of times this couple
appears on the greatest main interval  $[p+q+2,\,\delta]$ or
$[p+q+4,\,\beta]$ built from this value of $\delta$ or $\beta$.
\item The set of couples  $C$ with same value $p_{j}=k'$ ($j=2,\,3$)
constitutes a class of couples or more simply a class $C_{k'}$. The
population of a class is the number of couples belonging to this
class. \item The set of couples  $C$ with the same cumulative
multiplicity on $IS$ constitutes a class of multiplicity. We denote
by $\tilde{\Lambda}_{\mu}$ its population. \end{itemize}

\subsubsection{Properties}\label{prop}
\begin{itemize}
\item By construction, to a given value of $q_{3}$ corresponds $q_{3}-1$
$3$-couples.
\item A $2$-couple $(k',\,q_{2})$ or a $3$-couple $(k',\,q_{3},\,\gamma)$ , present, appears once and only one on an secondary interval:
their multiplicity is thus equal to 1.
\item The multiplicity of an absent couple on a main or secondary interval is equal to zero.
\item In the case of a resonance $p:q$, the cumulative multiplicity of a couple of integers  $C$ may take all integer values from $0$ to
$p+q$. This last value is the maximal cumulative multiplicity of the couple. This result is easily established from Eq. (\ref{eq.15}). Thus one deduces that the cumulative multiplicity of an absent couple is either zero or maximal. We will call switch-off couple, an absent couple of maximal cumulative multiplicity.
\end{itemize}

\subsubsection{Calculation of a cumulative multiplicity}\label{calcul}
By the same way, we calculate the cumulative multiplicity of a
$2$-couple $(k',\,q_{2})$ or a  $3$-couple $(k',\,q_{3},\,\gamma)$.
Also, the couples of integers belonging to a class $C_{k'}$ $(k',
\,q_{2})$ or $(k', \,q_{3},\,\gamma)$ do not still appear in the
counting and are absent (i.e. cumulative multiplicity equal to zero)
if $\delta <k'(p+q)+2q_{j}$ ($j=2,\,3$). To the contrary, for
$\delta>(k'+1)(p+q) + 2q_{j}-1$, they do no more appear: these
couples are switch-off; their cumulative multiplicity is maximal.
The only main intervals where the couples of integers are present
are of the form:
\begin{eqnarray}\label{eq.15}
[k'(p+q)+2q_{j},\,(k'+1)(p+q) + 2q_{j}-1].
\end{eqnarray}
For a given value of $N=N_{1}$ belonging to the interval given by
$(\ref{eq.15})$, the multiplicity of the present couple $(k', \,q_{2})$
or  $(k',\,q_{3},\,\gamma)$ on the main interval
$[k'(p+q)+2q_{j},\,N_{1}]$ may be calculated by:
\begin{eqnarray}\label{eq.16}
\mu = N_{1} - N_{app}+1\,,
\end{eqnarray}
with $N_{app}=k'(p+q)+2q_{j}$ the value of $N$ from which this couple appears on this main interval.
Eq. (\ref{eq.16})
gives also the cumulative multiplicity of the couple
present on the secondary interval $IS_{N_{1}}$. Indeed, from the definition, the cumulative multiplicity of the couple $(k', \,q_{2})$ or
$(k',\,q_{3},\,\gamma)$ on this secondary interval is its multiplicity on the largest main interval, here
$[p+q+2,\,N_{1}]$. But, $[p+q+2,\,N_{1}] = [p+q+2,\,N_{app}-1]
\bigcup [N_{app},\,N_{1}]$, thus from (\ref{eq.15}), on the main interval $[p+q+2,\,N_{app}-1]$, this couple is absent:
its multiplicity is equal to zero. By contrast, it is present on each of the $N_{1}-N_{app}+1$ secondary intervals associated
respectively to the values : $N_{app}$, ..., $N_{1}$. Its
multiplicity on $[N_{app},\,N_{1}]$ is equal to $N_{1}-N_{app}+1$.

\subsection{Theorems of the $2$-monomials counting}\label{Th2.mon}
\subsubsection{Pointing the problem} The three following theorems
about the counting give the number $\Delta_{1}$ of independent
monomials
${\sigma_{m}^{E(\frac{\delta-2q_{2}}{p+q})}}{\sigma_{\ell}^{q_{2}}}$
($m=-1,\,0$ and $\ell=1,\,2$) present in a sum $S_{m,\,\ell}^{(2)}$
but also the number $\tilde{\alpha}_{1}$ of $2$-monomials of degree
$N$. Without limiting the generality of the problem, we will write
$N=k'(p+q)+2+i$ with $k'$ a positive integer and $i$ an integer such
that $i\in [0,\,...,\,p+q-1]$.  {\textit{A posteriori}} we have to
distinguish trois cases in our study according the parity  of $p+q$
and of $k'$: $p+q$ even whatever the parity of $k'$; then for $p+q$
odd, to study the cases when $k'$ is even then odd. Demonstration of
these theorems is given in Appendix.

\begin{thm}
If $p+q$ is even and $k'$ an integer $\geq 1$:
\begin{equation}\label{theor.2a}
\Delta_{1} = k'[1+ E(\frac{i}{2})+\frac{(k'-1)(p+q)}{4}].
\end{equation}
\end{thm}

\begin{thm}
If $p+q$ odd and $k'$ an even integer $\geq 2$:
\begin{equation}\label{theor.2b}
\Delta_{1} = k'[\frac{(k'-1)(p+q)+2i+3}{4}].
\end{equation}
\end{thm}

\begin{thm}
If $p+q$ odd and $k'$ an odd integer $\geq 1$:
\begin{equation}\label{theor.2c}
\Delta_{1} = 1+ E(\frac{i}{2}) + (k'-1)[\frac{k'(p+q)+2i+3}{4}].
\end{equation}
\end{thm}

\begin{table}[h]\label{table.1.1}
\begin{center}
\begin{tabular}[h]{|c||c|c||c|c||c|c||c|c|} \hline
$N$ & $p+q$ & $\Delta_{1}$ & $p+q$ & $\Delta_{1}$ & $p+q$ & $\Delta_{1}$ & $p+q$ & $\Delta_{1}$ \\
\hline
 4  &   2   &     1        & 3     &        0      & 4    &       0      &   5   &     0         \\ \hline
 5  &   2   &     1        & 3     &            1  & 4    &       0      &   5   &     0         \\ \hline
 6  &   2   &     3        & 3     &            1  & 4    &       1      &   5   &     0         \\ \hline
 7  &   2   &     3        & 3     &            2  & 4    &       1      &   5   &     1         \\ \hline
 8  &   2   &     6        & 3     &            3  & 4    &       2      &   5   &     1         \\ \hline
 9  &   2   &     6        & 3     &            4  & 4    &       2      &   5   &     2         \\ \hline
 10  &   2   &     10      & 3     &            5  & 4    &       4      &   5   &     2        \\ \hline
 11  &   2   &     10      & 3     &            7  & 4    &       4      &   5   &     3         \\ \hline
 12  &   2   &     15      & 3     &            8  & 4    &       6      &   5   &     4         \\ \hline
 13  &   2   &     15      & 3     &            10  & 4    &      6      &   5   &     5         \\ \hline
 14  &   2   &     21       & 3     &           12  & 4    &      9      &   5   &     6         \\ \hline
 15  &   2   &     21       & 3     &           14  & 4    &      9      &   5   &     7         \\ \hline
 16  &   2   &     28       & 3     &           16  & 4    &     12      &   5   &     8           \\ \hline
 17  &   2   &     28       & 3     &           19  & 4    &     12      &   5   &     10            \\ \hline
 18  &   2   &     36       & 3     &           21  & 4    &     16      &   5   &     11          \\ \hline
\end{tabular} \caption{Counting of the $2$-monomials present in a
sum $S_{m,\,\ell}^{(2)}$ ($m=-1,\,0$; $\ell=1,\,2$) from
(\ref{eq.7}) for $4\leq N \leq 18$ and $2\leq p+q \leq 5$.}
\end{center} \end{table}

\subsection{Theorems of the $3$-monomials counting}\label{Th3.mon}
\subsubsection{Pointing the problem}
In this section, we give the theorems of the counting of the
$3$-monomials ${\sigma_{m}^{E(\frac{\beta-2q_{3}}{p+q})}}
{\sigma_{1}^{\gamma}{\sigma_{2}^{q_{3}-\gamma}}}$ ($m=-1,\,0$),
$\Delta_{2}$ in number,  appearing in a sum $S_{m}^{(3)}$ of
(\ref{eq.7}), but also the number  $\tilde{\alpha}_{2}$ of
$3$-monomials of degree $N$. Without limiting the generality of the
problem, we will write $N=k'(p+q)+4+i$ with $k'$ a positive integer
and $i$ an integer such that $i\in [0,\,...,\,p+q-1]$. As for the
$2$-monomials, three cases should be distinguished in our study
according to the parity of $p+q$ and of $k'$. In what follows, we
denote $\epsilon = i - 2E(\frac{i}{2})$: $\epsilon=0$ if $i$ even
and $\epsilon=1$ if $i$ odd.

\begin{thm}
If $p+q$ even and $k'$ an integer $\geq 1$:
\begin{eqnarray}\label{theor.3a}
\Delta_{2} &=& \frac{[E(\frac{i}{2})+1][E(\frac{i}{2})+2]}{2} \notag \\
&+& \frac{k'(k'-1)(p+q)}{48}[(2k'-1)(p+q)+6(i+3-\epsilon)] \notag \\
&+& \frac{(k'-1)}{8}[i(i+6)-4\epsilon E(\frac{i}{2})-7\epsilon +8].
\end{eqnarray}
\end{thm}

\begin{thm}
If $p+q$ odd and $k'$ an even integer $\geq 2$:
\begin{eqnarray}\label{theor.3b}
\Delta_{2} &=&
\frac{[E(\frac{i}{2})+1][E(\frac{i}{2})+2]}{2} \notag \\ &+&
\frac{k'(k'-1)(p+q)}{48}[(2k'-1)(p+q)+3(2i+5)] \notag \\ &+&
\frac{(k'-1)}{8}[i(i+6)-4\epsilon E(\frac{i}{2})-7\epsilon +8]
\notag \\ &+&
\frac{k'(2\epsilon-1)}{16}[4E(\frac{i}{2})+(p+q)+5+2\epsilon].
\end{eqnarray}
\end{thm}

\begin{thm}
If $p+q$ odd and $k'$ an odd integer $\geq 1$:
\begin{eqnarray}\label{theor.3c}
\Delta_{2} &=& \frac{[E(\frac{i}{2})+1][E(\frac{i}{2})+2]}{2} \notag \\
&+& \frac{k'(k'-1)(p+q)}{48}[(2k'-1)(p+q)+3(2i+5)] \notag \\
\notag &+& \frac{(k'-1)}{8}[i(i+6)-4\epsilon
E(\frac{i}{2})-7\epsilon +8] \notag \\
&+&
\frac{(k'-1)(2\epsilon-1)}{16}[4E(\frac{i}{2})-(p+q)+5+2\epsilon].
\end{eqnarray}
\end{thm}

\begin{table}[h]\label{table.2.1}
\begin{center}
\begin{tabular}[h]{|c||c|c||c|c||c|c||c|c|} \hline
$N$ & $p+q$ & $\Delta_{2}$ & $p+q$ & $\Delta_{2}$ & $p+q$ & $\Delta_{2}$ & $p+q$ & $\Delta_{2}$ \\
\hline
 6  &   2   &     1        & 3     &        0      & 4    &       0      &   5   &     0         \\ \hline
 7  &   2   &     1        & 3     &            1  & 4    &       0      &   5   &     0         \\ \hline
 8  &   2   &     4        & 3     &            1  & 4    &       1      &   5   &     0         \\ \hline
 9  &   2   &     4        & 3     &            3  & 4    &       1      &   5   &     1         \\ \hline
 10  &   2   &     10      & 3     &            4  & 4    &       3      &   5   &     1        \\ \hline
 11  &   2   &     10      & 3     &            7  & 4    &       3      &   5   &     3         \\ \hline
 12  &   2   &     20      & 3     &            9  & 4    &       7      &   5   &     3         \\ \hline
 13  &   2   &     20      & 3     &            14  & 4    &      7      &   5   &     6         \\ \hline
 14  &   2   &     35       & 3     &           17  & 4    &      13      &   5   &     7         \\ \hline
 15  &   2   &     35       & 3     &           24  & 4    &      13      &   5   &     11         \\ \hline
 16  &   2   &     56       & 3     &           29  & 4    &     22      &   5   &     13           \\ \hline
 17  &   2   &     56       & 3     &           38  & 4    &     22      &   5   &     18            \\ \hline
 18  &   2   &     84       & 3     &           45  & 4    &     34      &   5   &     21          \\ \hline
\end{tabular} \caption{ $3$-monomials counting present in a sum
$S_{m}^{(3)}$ ($m=-1,\,0$) of (\ref{eq.7}) for $6\leq N \leq 18$ and
$2\leq p+q \leq 5$.} \end{center} \end{table}

\begin{thm}
The normalized Hamiltonian $\mathcal{K}$ given by (\ref{eq.7}) with
a $p:q$ resonance between its two oscillators, is described by
$N_{coef}$ coefficients, $N_{op}$ independent monomials among which
$N_{c}$ are coupling monomials and given by:
\begin{eqnarray}
N_{coef} &=& \frac{Q_{0}(Q_{0}+3)}{2}+
E(\frac{N}{p+q})+2\Delta_{1}+\Delta_{2}, \label{coef}
\\
N_{op} &=& \frac{Q_{0}(Q_{0}+3)}{2}+2E(\frac{N}{p+q})+4\Delta_{1}+ 2\Delta_{2}, \label{op}\\
N_{c} &=& 2E(\frac{N}{p+q})+4\Delta_{1}+2\Delta_{2} \label{c}.
\end{eqnarray}

$\Delta_{1}$ and $\Delta_{2}$ are given by the counting theorems
(Theorems $1$ to $6$).
\end{thm}

\begin{table}[h]\label{table.3.1}
\begin{center}
\begin{tabular}[h]{|c|c|c|c|} \hline
$p+q$ & $N_{coef}$ & $N_{op}$ & $N_{c}$ \\ \hline

2 & 55 & 90 & 70
\\ \hline
3 & 37 & 54 & 34 \\ \hline

4 & 33 & 46 & 26 \\ \hline

5 & 27 & 34 & 14 \\ \hline

\end{tabular} \caption{Counting of the coefficients, monomials and
independent coupling monomials in (\ref{eq.7}) for $N = 10$ and
$2\leq p+q \leq 5$.} \end{center} \end{table}

\subsection{The general case}
\subsubsection{Construction of the normalized Hamiltonian}
We consider a Hamiltonian system described by  $n\geq 3$ oscillators among which
 oscillators "1" and "2" are in $p:q$ resonance. Hamiltonian $\mathcal{K}$ is supposed to be invariant under TRO. The quadratic part of
$\mathcal{K}$ is : $\mathcal{H}_{0} =
-\imath\sum_{k=1}^{n}\omega_{k}\sigma_{k}$. The $n$ quantities
$\omega_{k}$ are characteristic  pulsations of the oscillators.
The Hamiltonian can be expressed as a function of the generators of the Hilbert basis: $\mathcal{K} = \mathcal{H}_{0} +
f(\sigma_{-1},\,\sigma_{0},\,\sigma_{1},\,..., \, \sigma_{n})$.
If we write $\mathcal{K}$ as a polynomial development of the generators of the Hilbert basis  until the order $N\geq p+q+4$, it reads:

\begin{eqnarray*} \mathcal{K} =  \mathcal{H}_{0} +
\sum_{k=1}^{n}\sum_{q_{0}=2}^{Q_{0}}
{\alpha_{q_{0}}^{k}}{\sigma_{k}^{q_{0}}} + \end{eqnarray*}
\begin{eqnarray*} \sum_{\ell=2}^{n}\sum_{r=2}^{Q_{0}}\sum_{1\leq
i_{1} < i_{2} < \ldots < i_{\ell} \leq n}\sum_{r_{i_{1}} \geq
1,\,\ldots,\,r_{i_{\ell}}\geq1}^{r_{i_{1}}+\ldots +r_{i_{\ell}} = r}
{\alpha_{r_{i_{1}},\,...,\,r_{i_{\ell}}}^{i_{1},\,...,\,i_{\ell}}}
{\sigma_{i_{1}}^{r_{i_{1}}}}\ldots{\sigma_{i_{\ell}}^{r_{i_{\ell}}}}
\end{eqnarray*} \begin{eqnarray}\label{eq.100} &+&
\sum_{q_{1}=1}^{Q_{1}}
{\alpha_{q_{1}}^{0}}({\sigma_{-1}^{q_{1}}}+{\sigma_{0}^{q_{1}}})
\notag \\ & + &
\sum_{k=1}^{n}\sum_{\delta=p+q+2}^{N}\sum_{q_{2}=1}^{Q_{2}}
{\alpha_{E(\frac{\delta-2q_{2}}{p+q}),\,q_{2}}^{0,\,k}}\biggl({\sigma_{-1}^{E(\frac{\delta-2q_{2}}{p+q})}}
\notag \\ & + &
{\sigma_{0}^{E(\frac{\delta-2q_{2}}{p+q})}}\biggr){\sigma_{k}^{q_{2}}}
\notag \\ & + &
\sum_{i=1}^{n-1}\sum_{j=i+1}^{n}\sum_{\beta=p+q+4}^{N}\sum_{q_{3}=2}^{Q_{3}}\sum_{\gamma=1}^{q_{3}-1}
{\alpha_{E(\frac{\beta-2q_{3}}{p+q}),\,\gamma,\,q_{3}-\gamma}^{0,\,i,\,j}}
\notag \end{eqnarray} \begin{eqnarray} &&
\biggl({\sigma_{-1}^{E(\frac{\beta-2q_{3}}{p+q})}} +
{\sigma_{0}^{E(\frac{\beta-2q_{3}}{p+q})}}\biggr)
{\sigma_{i}^{\gamma}{\sigma_{j}^{q_{3}-\gamma}}}. \end{eqnarray} In
Eq.(\ref{eq.100}), all the coefficients are purely imaginary;
$i_{1}$, ..., $i_{\ell}$ ($2 \leq \ell \leq n$) are positive
integers satisfying the partial order: $1\leq i_{1} < i_{2} < \ldots
< i_{\ell} \leq n$; $r_{i_{1}}$, ..., $r_{i_{\ell}}$ are positive
integers satisfying the relation $r_{i_{1}}+...+r_{i_{\ell}} = r$,
with $r$ an integer between $2$ and $Q_{0}$; $Q_{0}=E(\frac{N}{2})$,
$Q_{1}=E(\frac{N}{p+q})$, $Q_{2}= E(\frac{\delta -(p+q)}{2})$ and
$Q_{3}= E(\frac{\beta -(p+q)}{2})$. \\ $\mathcal{K}$ involves two
contributions: a first one corresponding to a Dunham development on
the basis of the generators \cite{SLJ2015} and a second contribution
corresponding to a polynomial expansion of the coupling terms. \\
Furthermore, using the same method as here-before in \ref{ind}, one
obtains that the different coefficients involved in the normalized
Hamiltonian $\mathcal{K}$ given by Eq. (\ref{eq.100}) are
independent.

\begin{thm}\label{Th.5}
$\mathcal{K}$ is described by $N_{coef}$ coefficients (among which
$\frac{N_{c}}{2}$ coupling coefficients), $N_{op}$ independent
monomials whose $N_{c}$ are coupling monomials, satisfying the
following equations:
\begin{eqnarray}
N_{coef} &=& \Lambda+
E(\frac{N}{p+q})+n\Delta_{1}+\frac{n(n-1)}{2}\Delta_{2},
\label{coef2}
\\
N_{op} &=& \Lambda+2E(\frac{N}{p+q})+2n\Delta_{1} + n(n-1)\Delta_{2}, \label{op2}\\
N_{c} &=& 2E(\frac{N}{p+q})+2n\Delta_{1}+n(n-1)\Delta_{2}
\label{c2}.
\end{eqnarray}

$\Delta_{1}$ and $\Delta_{2}$ are given by the counting theorems
(Theorems $1$ to $6$) and $\Lambda =
\sum_{\lambda=1}^{min(n,\,Q_{0})}{C_{n}^{\lambda}}{C_{Q_{0}}^{\lambda}}$
(\cite{SLJ2015}).
\end{thm}

\section{Applications}\label{Appli}
\subsection{The molecule of ClOH}
\subsubsection{Conventions of notation}
ClOH is a non linear triatomic molecule with 3 vibrational degrees
of freedom.  ($n=3$). In the local modes representation, we attach a
stretching oscillator to each of the bonds Cl-O (oscillator "1") and
O-H (oscillator "3") and a bending oscillator (oscillator "2") to
the angle between these bonds.

\subsubsection{Quantum vibrational Hamiltonian} The classical
relations between dimensionless variables $\{{\underline{z}}_{j},
\,{\underline{z}}_{k}^{\ast}\} = -i\delta_{jk}$, take now the
following form: $\frac{1}{i}[a_{j}, a_{k}^{+}]$ $= -i\delta_{jk}$,
that is $1\leq j,\,k \leq 3$, $[a_{j}, a_{k}^{+}] = \delta_{jk}$.
These operators satisfy the Bose commutation relations and are
defined as the Boson creation operator $a_{k}^{+}$ and Boson
annihilation operator $a_{j}$. \\ Generators $\sigma_{k}$ and
Hamiltonian function $\mathcal{K}$ are respectively replaced by the
number operators
 $\hat{N}_{k}= a_{k}^{+}a_{k}$, which physically express the number of quanta of excitation of each oscillator
$k$, and the Hamiltonian operator $\hat{K}$. By convention, for the expression of powers of number operators, we adopt the following form (\cite{JOY1999}):
\begin{eqnarray}
{\hat{N}_{i_{1}}}^{r_{1}}\ldots
{\hat{N}_{i_{\ell}}}^{r_{\ell}}=\underbrace{{a_{i_{1}}^{+}}a_{i_{1}}\ldots
{a_{i_{1}}^{+}}a_{i_{1}}}_{r_{1}\,\,\text{times}}\ldots
\underbrace{{a_{i_{\ell}}^{+}}a_{i_{\ell}}\ldots
{a_{i_{\ell}}^{+}}a_{i_{\ell}}}_{r_{\ell}\,\,\text{times}};
\end{eqnarray}
but we will write under normal form, as a function of the Bose operators, all the coupling operators:
$\hat{\sigma}_{-1}+{\hat{\sigma_{0}}} =
{a_{1}^{+}}^{p}{a_{2}}^{q}+{a_{2}^{+}}^{q}{a_{1}}^{p}$,
\begin{eqnarray}\label{eq.101}
\hat{\sigma}_{-1}^{q_{1}}+{\hat{\sigma_{0}}}^{q_{1}} &=&
\underbrace{{a_{1}^{+}}^{p}\ldots{a_{1}^{+}}^{p}}_{q_{1}\,\,\text{times}}
\underbrace{{a_{2}}^{q}\ldots{a_{2}}^{q}}_{q_{1}\,\,\text{times}}
\notag
\\
&+&\underbrace{{a_{2}^{+}}^{q}\ldots{a_{2}^{+}}^{q}}_{q_{1}\,\,\text{times}}
\underbrace{{a_{1}}^{p}\ldots{a_{1}}^{p}}_{q_{1}\,\,\text{times}}.
\end{eqnarray}
On a more compact way, Eq. (\ref{eq.101}) may be rewritten:
\begin{eqnarray}\label{eq.102}
\hat{\sigma}_{-1}^{q_{1}}+{\hat{\sigma_{0}}}^{q_{1}} =
({a_{1}^{+}}^{p}{a_{2}}^{q})^{q_{1}}+({a_{2}^{+}}^{q}{a_{1}}^{p})^{q_{1}}.
\end{eqnarray}
Similarly, the treatment of others coupling operators in
$\hat{K}$ gives:
\begin{eqnarray}
(\hat{\sigma}_{-1}^{p_{2}} + {\hat{\sigma_{0}}}^{p_{2}})
{\hat{N}_{k}}^{q_{2}} =\notag \\
\biggl(({a_{1}^{+}}^{p}{a_{2}}^{q})^{p_{2}}
+({a_{2}^{+}}^{q}{a_{1}}^{p})^{p_{2}}\biggr)
\underbrace{{a_{k}^{+}}a_{k}...{a_{k}^{+}}a_{k}}_{q_{2}\,\,\text{times}}.
\end{eqnarray}
\begin{eqnarray}
(\hat{\sigma}_{-1}^{p_{3}} + {\hat{\sigma_{0}}}^{p_{3}})
{\hat{N}_{i}}^{\gamma}{\hat{N}_{j}}^{q_{3}-\gamma} =\notag \\
\biggl(({a_{1}^{+}}^{p}{a_{2}}^{q})^{p_{3}}+({a_{2}^{+}}^{q}{a_{1}}^{p})^{p_{3}}\biggr)
\underbrace{{a_{i}^{+}}a_{i}...{a_{i}^{+}}a_{i}}_{\gamma\,\,\text{times}}
\underbrace{{a_{j}^{+}}a_{j}...{a_{j}^{+}}a_{j}}_{q_{3}-\gamma\,\,\text{times}}.
\end{eqnarray}
Hamiltonian $\hat{K}$ is Hermitian and the $N_{coef}$
coefficients are real.

\subsubsection{Eigen basis of the Hamiltonian $\hat{H_{0}}$} The
eigenstates of $\hat{H_{0}}$ are generated from the vacuum state
with the relation ($n_{1}$, $n_{2}$, $n_{3}$ are integers):
\begin{equation}\label{eq.4.1.1} \left| \begin{array}{c} n_{1},
n_{2},n_{3} \\ \end{array} \right >
=\frac{1}{\sqrt{n_{1}!n_{2}!n_{3}!}}
{a_{1}^{+}}^{n_{1}}{a_{2}^{+}}^{n_{2}}{a_{3}^{+}}^{n_{3}} \left|
\begin{array}{c} 0,0,0 \\ \end{array} \right >\,. \end{equation}

>From Theorem $8$,  the quantum vibrational Hamiltonian, developed
until the order $N=10$ ($Q_{0}=5$), is described by 115 operators,
60 of which are coupling operators and  85 coefficients; so, we
write (with the usual convention that $\hbar$ is equal to 1):
\begin{eqnarray} \hat{K} &=&
\biggl(\omega_{1}{\hat{N}_{1}}+\omega_{2}{\hat{N}_{2}}+\omega_{3}{\hat{N}_{3}}
\notag \\ &+&
\alpha_{2}^{1}{\hat{N}_{1}}^{2}+\alpha_{3}^{1}{\hat{N}_{1}}^{3}+\alpha_{4}^{1}{\hat{N}_{1}}^{4}+\alpha_{5}^{1}{\hat{N}_{1}}^{5}
\notag \\ &+&
\alpha_{2}^{2}{\hat{N}_{2}}^{2}+\alpha_{3}^{2}{\hat{N}_{2}}^{3}+\alpha_{4}^{2}{\hat{N}_{2}}^{4}+\alpha_{5}^{2}{\hat{N}_{2}}^{5}
\notag \\ &+&
\alpha_{2}^{3}{\hat{N}_{3}}^{2}+\alpha_{3}^{3}{\hat{N}_{3}}^{3}+\alpha_{4}^{3}{\hat{N}_{3}}^{4}+\alpha_{5}^{3}{\hat{N}_{3}}^{5}
\notag \\ &+&
\alpha_{1,1}^{1,2}{\hat{N}_{1}}{\hat{N}_{2}}+\alpha_{1,2}^{1,2}{\hat{N}_{1}}{\hat{N}_{2}}^{2}+\alpha_{2,1}^{1,2}{\hat{N}_{1}}^{2}{\hat{N}_{2}}
\notag \\ &+&
\alpha_{1,3}^{1,2}{\hat{N}_{1}}{\hat{N}_{2}}^{3}+\alpha_{2,2}^{1,2}{\hat{N}_{1}}^{2}{\hat{N}_{2}}^{2}+\alpha_{3,1}^{1,2}{\hat{N}_{1}}^{3}{\hat{N}_{2}}
\notag \\ &+&
\alpha_{1,4}^{1,2}{\hat{N}_{1}}{\hat{N}_{2}}^{4}+\alpha_{2,3}^{1,2}{\hat{N}_{1}}^{2}{\hat{N}_{2}}^{3}+\alpha_{3,2}^{1,2}{\hat{N}_{1}}^{3}{\hat{N}_{2}}^{2}
+\alpha_{4,1}^{1,2}{\hat{N}_{1}}^{4}{\hat{N}_{2}} \notag \\ &+&
\alpha_{1,1}^{1,3}{\hat{N}_{1}}{\hat{N}_{3}}+\alpha_{1,2}^{1,3}{\hat{N}_{1}}{\hat{N}_{3}}^{2}+\alpha_{2,1}^{1,3}{\hat{N}_{1}}^{2}{\hat{N}_{3}}
\notag \\ &+&
\alpha_{1,3}^{1,3}{\hat{N}_{1}}{\hat{N}_{3}}^{3}+\alpha_{2,2}^{1,3}{\hat{N}_{1}}^{2}{\hat{N}_{3}}^{2}+\alpha_{3,1}^{1,3}{\hat{N}_{1}}^{3}{\hat{N}_{3}}
\notag \\ &+&
\alpha_{1,4}^{1,3}{\hat{N}_{1}}{\hat{N}_{3}}^{4}+\alpha_{2,3}^{1,3}{\hat{N}_{1}}^{2}{\hat{N}_{3}}^{3}+\alpha_{3,2}^{1,3}{\hat{N}_{1}}^{3}{\hat{N}_{3}}^{2}
+\alpha_{4,1}^{1,3}{\hat{N}_{1}}^{4}{\hat{N}_{3}} \notag \\ &+&
\alpha_{1,1}^{2,3}{\hat{N}_{2}}{\hat{N}_{3}}+\alpha_{1,2}^{2,3}{\hat{N}_{2}}{\hat{N}_{3}}^{2}+\alpha_{2,1}^{2,3}{\hat{N}_{2}}^{2}{\hat{N}_{3}}
\notag \\ &+&
\alpha_{1,3}^{2,3}{\hat{N}_{2}}{\hat{N}_{3}}^{3}+\alpha_{2,2}^{2,3}{\hat{N}_{2}}^{2}{\hat{N}_{3}}^{2}+\alpha_{3,1}^{2,3}{\hat{N}_{2}}^{3}{\hat{N}_{3}}
\notag \\ &+&
\alpha_{1,4}^{2,3}{\hat{N}_{2}}{\hat{N}_{3}}^{4}+\alpha_{2,2}^{2,3}{\hat{N}_{2}}^{2}{\hat{N}_{3}}^{2}+\alpha_{3,2}^{2,3}{\hat{N}_{2}}^{3}{\hat{N}_{3}}^{2}
+\alpha_{4,1}^{2,3}{\hat{N}_{2}}^{4}{\hat{N}_{3}} \notag \\ &+&
\alpha_{1,1,1}^{1,2,3}{\hat{N}_{1}}{\hat{N}_{2}}{\hat{N}_{3}} \notag
\\ &+&
\alpha_{1,1,2}^{1,2,3}{\hat{N}_{1}}{\hat{N}_{2}}{\hat{N}_{3}}^{2}
+\alpha_{1,2,1}^{1,2,3}{\hat{N}_{1}}{\hat{N}_{2}}^{2}{\hat{N}_{3}}+\alpha_{2,1,1}^{1,2,3}{\hat{N}_{1}}^{2}{\hat{N}_{2}}{\hat{N}_{3}}
\notag \\ &+&
\alpha_{1,1,3}^{1,2,3}{\hat{N}_{1}}{\hat{N}_{2}}{\hat{N}_{3}}^{3}
+\alpha_{1,3,1}^{1,2,3}{\hat{N}_{1}}{\hat{N}_{2}}^{3}{\hat{N}_{3}}+\alpha_{3,1,1}^{1,2,3}{\hat{N}_{1}}^{3}{\hat{N}_{2}}{\hat{N}_{3}}
\notag \\ &+&
\alpha_{1,2,2}^{1,2,3}{\hat{N}_{1}}{\hat{N}_{2}}^{2}{\hat{N}_{3}}^{2}
+\alpha_{2,1,2}^{1,2,3}{\hat{N}_{1}}^{2}{\hat{N}_{2}}{\hat{N}_{3}}^{2}+\alpha_{2,2,1}^{1,2,3}{\hat{N}_{1}}^{2}{\hat{N}_{2}}^{2}{\hat{N}_{3}}
\notag \\ &+&
\alpha_{1}^{0}({a_{2}^{+}}{a_{1}}^{2}+{a_{1}^{+}}^{2}{a_{2}})
+\alpha_{2}^{0}({a_{2}^{+}}^{2}{a_{1}}^{4}
+{a_{1}^{+}}^{4}{a_{2}}^{2}) \notag \\ &+&
\alpha_{3}^{0}({a_{2}^{+}}^{3}{a_{1}}^{6}
+{a_{1}^{+}}^{6}{a_{2}}^{3}) \notag \\ &+&
\alpha_{1,1}^{0,1}({a_{2}^{+}}{a_{1}}^{2}+{a_{1}^{+}}^{2}{a_{2}}){\hat{N}_{1}}
+\alpha_{1,2}^{0,1}({a_{2}^{+}}{a_{1}}^{2}+{a_{1}^{+}}^{2}{a_{2}}){\hat{N}_{1}}^{2}
\notag \\ &+&
\alpha_{2,1}^{0,1}({a_{2}^{+}}^{2}{a_{1}}^{4}+{a_{1}^{+}}^{4}{a_{2}}^{2}){\hat{N}_{1}}
+\alpha_{1,3}^{0,1}({a_{2}^{+}}{a_{1}}^{2}+{a_{1}^{+}}^{2}{a_{2}}){\hat{N}_{1}}^{3}
\notag \\ &+&
\alpha_{2,2}^{0,1}({a_{2}^{+}}^{2}{a_{1}}^{4}+{a_{1}^{+}}^{4}{a_{2}}^{2}){\hat{N}_{1}}^{2}
\notag \\ &+&
\alpha_{1,1}^{0,2}({a_{2}^{+}}{a_{1}}^{2}+{a_{1}^{+}}^{2}{a_{2}}){\hat{N}_{2}}
+\alpha_{1,2}^{0,2}({a_{2}^{+}}{a_{1}}^{2}+{a_{1}^{+}}^{2}{a_{2}}){\hat{N}_{2}}^{2}
\notag \\ &+&
\alpha_{2,1}^{0,2}({a_{2}^{+}}^{2}{a_{1}}^{4}+{a_{1}^{+}}^{4}{a_{2}}^{2}){\hat{N}_{2}}
+\alpha_{1,3}^{0,2}({a_{2}^{+}}{a_{1}}^{2}+{a_{1}^{+}}^{2}{a_{2}}){\hat{N}_{2}}^{3}
\notag \\ &+&
\alpha_{2,2}^{0,2}({a_{2}^{+}}^{2}{a_{1}}^{4}+{a_{1}^{+}}^{4}{a_{2}}^{2}){\hat{N}_{2}}^{2}
\notag \\ &+&
\alpha_{1,1}^{0,3}({a_{2}^{+}}{a_{1}}^{2}+{a_{1}^{+}}^{2}{a_{2}}){\hat{N}_{3}}
+\alpha_{1,2}^{0,3}({a_{2}^{+}}{a_{1}}^{2}+{a_{1}^{+}}^{2}{a_{2}}){\hat{N}_{3}}^{2}
\notag \end{eqnarray} \begin{eqnarray}\label{big} &+&
\alpha_{2,1}^{0,3}({a_{2}^{+}}^{2}{a_{1}}^{4}+{a_{1}^{+}}^{4}{a_{2}}^{2}){\hat{N}_{3}}
+\alpha_{1,3}^{0,3}({a_{2}^{+}}{a_{1}}^{2}+{a_{1}^{+}}^{2}{a_{2}}){\hat{N}_{3}}^{3}
\notag \\ &+&
\alpha_{2,2}^{0,3}({a_{2}^{+}}^{2}{a_{1}}^{4}+{a_{1}^{+}}^{4}{a_{2}}^{2}){\hat{N}_{3}}^{2}
\notag \\ &+&
\alpha_{1,1,1}^{0,1,2}({a_{2}^{+}}{a_{1}}^{2}+{a_{1}^{+}}^{2}{a_{2}}){\hat{N}_{1}}{\hat{N}_{2}}
\notag \\
&+&\alpha_{1,1,1}^{0,1,3}({a_{2}^{+}}{a_{1}}^{2}+{a_{1}^{+}}^{2}{a_{2}}){\hat{N}_{1}}{\hat{N}_{3}}
\notag \\
&+&\alpha_{1,1,1}^{0,2,3}({a_{2}^{+}}{a_{1}}^{2}+{a_{1}^{+}}^{2}{a_{2}}){\hat{N}_{2}}{\hat{N}_{3}}
\notag \\
&+&\alpha_{1,1,2}^{0,1,2}({a_{2}^{+}}{a_{1}}^{2}+{a_{1}^{+}}^{2}{a_{2}}){\hat{N}_{1}}{\hat{N}_{2}}^{2}
\notag \\
&+&\alpha_{1,1,2}^{0,1,3}({a_{2}^{+}}{a_{1}}^{2}+{a_{1}^{+}}^{2}{a_{2}}){\hat{N}_{1}}{\hat{N}_{3}}^{2}
\notag \\
&+&\alpha_{1,1,2}^{0,2,3}({a_{2}^{+}}{a_{1}}^{2}+{a_{1}^{+}}^{2}{a_{2}}){\hat{N}_{2}}{\hat{N}_{3}}^{2}
\notag \\
&+&\alpha_{1,2,1}^{0,1,2}({a_{2}^{+}}{a_{1}}^{2}+{a_{1}^{+}}^{2}{a_{2}}){\hat{N}_{1}}^{2}{\hat{N}_{2}}
\notag \\
&+&\alpha_{1,2,1}^{0,1,3}({a_{2}^{+}}{a_{1}}^{2}+{a_{1}^{+}}^{2}{a_{2}}){\hat{N}_{1}}^{2}{\hat{N}_{3}}
\notag \\
&+&\alpha_{1,2,1}^{0,2,3}({a_{2}^{+}}{a_{1}}^{2}+{a_{1}^{+}}^{2}{a_{2}}){\hat{N}_{2}}^{2}{\hat{N}_{3}}
\notag \\
&+&\alpha_{2,1,1}^{0,1,2}({a_{2}^{+}}^{2}{a_{1}}^{4}+{a_{1}^{+}}^{4}{a_{2}}^{2}){\hat{N}_{1}}{\hat{N}_{2}}
\notag \\
&+&\alpha_{2,1,1}^{0,1,3}({a_{2}^{+}}^{2}{a_{1}}^{4}+{a_{1}^{+}}^{4}{a_{2}}^{2}){\hat{N}_{1}}{\hat{N}_{3}}
\notag \\
&+&\alpha_{2,1,1}^{0,2,3}({a_{2}^{+}}^{2}{a_{1}}^{4}+{a_{1}^{+}}^{4}{a_{2}}^{2}){\hat{N}_{2}}{\hat{N}_{3}}
\biggr). \end{eqnarray}

\subsubsection{Numerical simulations }\label{simul}

\begin{table}[h] \begin{center} \begin{eqnarray*}
\begin{array}{|c|c|c|}\hline N &  \Lambda & \text{coefficients in cm}^{-1} \\ \hline 2 & 3 & \omega_{1}= +753. 834, \,
\omega_{2}=+1\,258.914\\
 & & \omega_{3}= +3\,777.067 \\
\hline
4 & 9 & {\alpha_{2}^{1}} = -7.123, \,{\alpha_{2}^{2}}=+3.204,\,{\alpha_{2}^{3}}=-80.277 \\
  &   & \alpha_{1,1}^{1,2}=-10.637,\,\alpha_{1,1}^{1,3}=0,\, \alpha_{1,1}^{2,3}=-19.985\\
\hline
6 & 19 & {\alpha_{3}^{1}}= +0.0825, \, ,\,{\alpha_{3}^{2}}= 0,\, {\alpha_{3}^{3}}= -0.3619 \\
  &   & \alpha_{1,2}^{1,2}= -0.2503, \alpha_{1,2}^{1,3}=-0.0532,\,\alpha_{1,2}^{2,3}= -1.9534 \\
  &   & \alpha_{2,1}^{1,2}= -0.0802,\,\alpha_{2,1}^{1,3}= 0,\,  \alpha_{2,1}^{2,3}= 0  \\
  &   & \alpha_{1,1,1}^{1,2,3}=0 \\
\hline
8 & 34 & {\alpha_{4}^{1}}= -0.00171,\,{\alpha_{4}^{2}}= -0.04117, \, {\alpha_{4}^{3}}= 0   \\
   &   & \alpha_{3,1}^{1,2}=0,\, \alpha_{3,1}^{1,3}=0,\, \alpha_{3,1}^{2,3}=0 \\
   &   & \alpha_{2,2}^{1,2}= 0, \, \alpha_{2,2}^{1,3}=0,\, \alpha_{2,2}^{2,3}= -0.15070 \\
   &   & \alpha_{1,3}^{1,2}= -0.01229,\, \alpha_{1,3}^{1,3}=0,\, \alpha_{1,3}^{2,3}=
    +0.13189\\
  &   & \alpha_{1,1,2}^{1,2,3}= +0.02381,\, \alpha_{1,2,1}^{1,2,3}=0,\,\alpha_{2,1,1}^{1,2,3}=0 \\
 \hline
10 & 55 & {\alpha_{5}^{1}}= 0,\,{\alpha_{5}^{2}}=
+0.00151,\,{\alpha_{5}^{3}}= 0 \\
  &   & \,
  \alpha_{4,1}^{1,2}=0,\,\alpha_{4,1}^{1,3}=0,\,\alpha_{4,1}^{2,3}=0
  \\
   &   & \,
  \alpha_{3,2}^{1,2}=0,\,\alpha_{3,2}^{1,3}=0,\,\alpha_{3,2}^{2,3}=0
  \\
  &   & \,
  \alpha_{2,3}^{1,2}=0,\,\alpha_{2,3}^{1,3}=0,\,\alpha_{2,3}^{2,3}=-0.00066
  \\
   &   & \,
  \alpha_{1,4}^{1,2}= 0,\,\alpha_{1,4}^{1,3}=0,\,\alpha_{1,4}^{2,3}=0
  \\
\hline
\end{array}
\end{eqnarray*}
\caption{List of the Dunham coefficients given by
\cite{JJS1999}. For  a given order  $N$  ($4\leq N \leq 10$), a line contains the numbers of additional coefficients  to the
order $N-2$.}\label{table.2}
\end{center}
\end{table}

The vibrational structure of the ClOH molecule has been studied in
\cite{JJS1999} until almost the dissociation limit. For levels of
energy less than $70$ $\%$ of the dissociation limit, the authors
make a Dunham expression based on the number operators (See
\cite{SLJ2015}). But, for highly excited levels, due to the more and
more frequent accidentally couplings between levels energetically
close, Fermi resonance between  oscillators "1" and "2"
($\frac{\omega_{2}}{\omega_{1}}\approx 2$) has to be taken into
account. Quantum numbers  $n_{1}$ and $n_{2}$ are "no more good
quantum numbers", as refers in the literature, and they are replaced
by the polyad number $P=n_{1}+2n_{2}$ (See for instance
\cite{JS2002}). In \cite{JJS1999}, the authors determine $725$
levels of energy, which means to take into account up to $38$ quanta
of excitation exchanged between  oscillator "1" and  "2" ($P \leq
38$) and $7$ quanta for the oscillator "3", labeling of the states
being made with the polyad number  $[P,\,n_{3}]$. Furthermore, these
authors add a coupling operator
${\alpha'_{1}}^{0}({a_{2}^{+}}^{3}{a_{3}}+{a_{3}^{+}}{a_{2}}^{3})$
in the Hamiltonian $\hat{K}$ in order to describe the $3:1$
resonance between oscillators  "2" and "3"
($\frac{\omega_{3}}{\omega_{2}}\approx 3$). Our model predicts that
the Hamiltonian $\hat{K}$ has to be described by $86$ coefficients
(the $85$ coefficients of Eq. (\ref{big}) + ${\alpha'_{1}}^{0}$).
However the smallest rms value ($=5.29$ cm$^{-1}$) is obtained for a
fit with only $28$ coefficients different from zero, some
coefficients have been set at zero by a more or less arbitrary way.
Results are given in the tables \ref{table.2} and \ref{table.3}.

\begin{table}[h] \begin{center} \begin{eqnarray*}
\begin{array}{|c|c|c|}\hline N &  \frac{N_{c}}{2} &
\text{coefficients in cm}^{-1} \\ \hline 3 & 1 & {\alpha_{1}^{0}}= 0
\\ \hline 4 & 1 & ({\alpha'_{1}}^{0}=+0.19520)
\\ \hline 5 & 4 & \alpha_{1,1}^{0,1} = -0.24939,
\,\alpha_{1,1}^{0,2} = 0,\,\alpha_{1,1}^{0,3} = -0.76017  \\ \hline
6 & 5 & {\alpha_{2}^{0}}= 0 \\ \hline 7 & 11 & \alpha_{1,2}^{0,1} =
+0.00583,\,\alpha_{1,2}^{0,2} =0,\,\alpha_{1,2}^{0,3}= -0.01158 \\
  & & \alpha_{1,1,1}^{0,1,2}=+0.04075,\,\alpha_{1,1,1}^{0,1,3}=0,\,\alpha_{1,1,1}^{0,2,3}=
  0 \\
\hline
8 & 14  & \alpha_{2,1}^{0,1}=0,\,\alpha_{2,1}^{0,2}=0,\,\alpha_{2,1}^{0,3}=0 \\
\hline
9 & 24 & {\alpha_{3}^{0}}= 0,\,\alpha_{1,3}^{0,1}=0,\,\alpha_{1,3}^{0,2}=0,\,\alpha_{1,3}^{0,3}=0 \\
 & &
 \alpha_{1,1,2}^{0,1,2}=0,\,\alpha_{1,2,1}^{0,1,2}=0,\,\alpha_{1,1,2}^{0,1,3}=0,\,\alpha_{1,2,1}^{0,1,3}=0
 \\
 & & \alpha_{1,1,2}^{0,2,3}=0,\,\alpha_{1,2,1}^{0,2,3}=0 \\
\hline
10 & 30 & \alpha_{2,1,1}^{0,1,2}=0,\,\alpha_{2,1,1}^{0,1,3}=0,\,\alpha_{2,1,1}^{0,2,3}=0 \\
 & &
 \alpha_{2,2}^{0,1}=0,\,\alpha_{2,2}^{0,2}=0,\,\alpha_{2,2}^{0,3}=0
 \\
\hline
\end{array}
\end{eqnarray*}
\caption{List of the coupling coefficients given by
\cite{JJS1999}. For a given order $N$  ($4\leq N \leq 10$), a line contains the number of additional coefficients to the order $N-1$. The coupling coefficient of the $3:1$ resonance is in brackets and is not accounting for in the enumeration.}\label{table.3}
\end{center}
\end{table}

\section{Conclusion and perspectives}\label{Concl}
We have presented a method of construction of a vibrational normalized Hamiltonian, modeled by a set of
$n$ oscillators until a high order $N$. It allows to describe the highly excited vibrational levels in the case of a  $p:q$ resonance.
 We have also counted all the operators introduced in the Hamiltonian, in particular the coupling operators.
 This building method has been successfully applied to the ClOH molecule taking into account a $2:1$ resonance.
 To go further, in a strict manner, for molecular systems having at least 3 oscillators, we should add in the Hamiltonian polynomial expansion, $4$-monomials
 based on the generators of the invariant algebra (Eq.(\ref{eq.100})) as soon as $N \geq p+q+6$. If the method of construction is easily
 adaptable, the counting theorems for these monomials remain to be done.
\section{Acknowledgments}\label{Ack}
Authors thank Prof. Oleg N. Ulenikov and Elena S. Bekhtereva from
Tomsk Polytechnic University, for valuable discussions. Research
conducted in the scope of the project FTI 120.

\section{Appendix}
\subsection{$2$-monomials counting}
In this section, we present the demonstration of the three counting
theorems of the $2$-monomials. We denote the order $N$ of the
development of (\ref{eq.100}) under the form $N=k'(p+q)+2+i$ with
$k'$ and $i\in[0,\,...,\,p+q-1]$ two positive integers. Thus we are
working on the secondary interval $IS=[1,\,Q_{2}]_{N=k'(p+q)+2+i}$
with $Q_{2}=E(\frac{(k'-1)(p+q)+2+i}{2})$.

\subsubsection{The different classes of couples}
On $IS$, the different values taken by $q_{2}$ belong to different classes of couples: $C_{1},\,...,\,C_{k'-1}$ and
$C_{k'}$. We begin by searching the couples belonging to the class  $C_{k'}$ as well as the population  $\tilde{\Lambda}''$ of this class.
The couples $(k',\,q_{2})$ of this class are: $(k',\,1)$,
$(k',\,2)$, $...$, $(k',\tilde{q}_{2})$ with $\tilde{q}_{2}$ the highest possible value of $q_{2}$ such that $p_{2} =
E(\frac{k'(p+q)+2+i-2\tilde{q}_{2}}{p+q})=k'$, that is
$\tilde{q}_{2}=1+E(\frac{i}{2})$. the multiplicities of these
$1+E(\frac{i}{2})$ couples are easily calculated with the Eq.
(\ref{eq.15}) and (\ref{eq.16}). The cumulative multiplicities of the couples $(k',\,1)$, $(k',\,2)$, $...$, $(k',\tilde{q}_{2})$ on $IS$
are respectively equal to
$\mu=i+1,\,i-1,\,...,\,\,i+3-2q_{2},\,...,\,1+\epsilon$ with
$\epsilon=i-2E(\frac{i}{2})$. It is possible now to evaluate the numbers of couples of this class: the population of $C_{k'}$ is the sum of all the couples present in this class taking into account the cumulative  multiplicity of each of the couples on $IS$, which is equivalent to calculate the number of times a couple appears on the main interval
 $[p+q+2,\,k'(p+q)+2+i]$. We get:
\begin{eqnarray}\label{eq.17}
\tilde{\Lambda}'' &=& \sum_{q_{2}=1}^{\tilde{q}_{2}}(i+3-2q_{2}), \notag \\
&=& [E(\frac{i}{2})+1][i+1-E(\frac{i}{2})],
\end{eqnarray}
or accordingly to the parity of $i$:
\begin{equation}\label{eq.18}
\tilde{\Lambda}''=\frac{i^{2}}{4}+i+1 \,\,\text{($i$ even)},
\end{equation}
\begin{equation}\label{eq.19}
\tilde{\Lambda}''=\frac{i^{2}}{4}+i+\frac{3}{4}\,\,\text{($i$
odd)}.
\end{equation}
Couples of the others different classes of $IS$ break down as follows: $(k'-1, E(\frac{i}{2})+2)$, $(k'-1, E(\frac{i}{2})+3)$,
$...$, $(k'-1, E(\frac{p+q+i+2}{2}))$ for the class $C_{k'-1}$,
$(k'-2, E(\frac{p+q+i+4}{2}))$, $(k'-2, E(\frac{p+q+i+6}{2}))$,
$...$, $(k'-2, E(\frac{2(p+q)+i+2}{2}))$ for the class $C_{k'-2}$,
$...$, $(2, E(\frac{(k'-3)(p+q)+i+4}{2}))$, \\
$(2, E(\frac{(k'-3)(p+q)+i+6}{2}))$, $...$, $(2,
E(\frac{(k'-2)(p+q)+i+2}{2}))$ for the class $C_{2}$ and finally $(1,
E(\frac{(k'-2)(p+q)+i+4}{2}))$, \\
$(1, E(\frac{(k'-2)(p+q)+i+6}{2}))$, $...$, $(1,
Q_{2}=E(\frac{(k'-1)(p+q)+i+2}{2}))$ for the class $C_{1}$.

Calculation of the  cumulative multiplicity of the $2$-couples
implies to distinguish three cases: $p+q$ even (case A), $p+q$ odd
with  $k'$ even (case B) and $p+q$ odd with $k'$ odd (case C).
Results are given in the tables \ref{table.Appendix.1} to
\ref{table.Appendix.5}.

\begin{table}[H] \begin{center} \begin{tabular}[h]{|c|c|c|} \hline
class $C_{j}$ & $q_{2}$  &  $\mu$ \\ \hline $C_{k'-1}$ &
$E(\frac{i}{2})+2$ &  $p+q-1+\epsilon$         \\ $C_{k'-1}$ &
$E(\frac{i}{2})+3$ & $p+q-3+\epsilon$         \\ \vdots & \vdots &
\vdots \\ $C_{k'-1}$ & $E(\frac{i}{2})+\frac{p+q}{2}+1$ &
$1+\epsilon$ \\ \hline $C_{k'-2}$ &$E(\frac{i}{2})+\frac{p+q}{2}+2$
& $p+q-1+\epsilon$ \\ $C_{k'-2}$ &$E(\frac{i}{2})+\frac{p+q}{2}+3$ &
$p+q-3+\epsilon$ \\ \vdots & \vdots & \vdots \\ $C_{k'-2}$ &
$E(\frac{i}{2})+p+q+1$ & $1+\epsilon$ \\ \hline \vdots & \vdots &
\vdots \\ \hline $C_{2}$ & $E(\frac{i}{2})+\frac{(k'-3)(p+q)}{2}+2$
& $p+q-1+\epsilon$ \\ $C_{2}$ &
$E(\frac{i}{2})+\frac{(k'-3)(p+q)}{2}+3$ & $p+q-3+\epsilon$ \\
\vdots & \vdots & \vdots \\ $C_{2}$ &
$E(\frac{i}{2})+\frac{(k'-2)(p+q)}{2}+1$ & $1+\epsilon$ \\ \hline
$C_{1}$ & $E(\frac{i}{2})+\frac{(k'-2)(p+q)}{2}+2$ &
$p+q-1+\epsilon$ \\ $C_{1}$ &
$E(\frac{i}{2})+\frac{(k'-2)(p+q)}{2}+3$ & $p+q-3+\epsilon$ \\
\vdots & \vdots & \vdots \\ $C_{1}$ &
$E(\frac{i}{2})+\frac{(k'-1)(p+q)}{2}+1$ & $1+\epsilon$ \\ \hline
\end{tabular} \caption{Case $A$. Table giving the different classes
of couples with the cumulative multiplicities . If $i$ even
($\epsilon=0$), all the cumulative multiplicity are odd; they are
all even if $i$ is odd ($\epsilon=1$).}\label{table.Appendix.1}
\end{center} \end{table}

\subsubsection{Case $A$} We denote by $\tilde{\Lambda}'$ the
population of the classes $C_{k'-1}$ to $C_{1}$; on each of these
classes, each cumulative multiplicity ($\frac{p+q}{2}$ in number)
$p+q-1+\epsilon$, $p+q-3+\epsilon$, $...$, $1+\epsilon$ appears only
one time; we may write: \begin{eqnarray}\label{eq.20}
\tilde{\Lambda}' &=&
(k'-1)\sum_{j=0}^{\frac{p+q-2}{2}}(2j+1+\epsilon), \notag \\ & = &
\frac{(k'-1)(p+q)}{2}[\frac{p+q}{2}+\epsilon]. \end{eqnarray} The
population $\tilde{\Lambda}$ of all the classes $C_{j}$
($j=1,\,...,\,k'$) is the sum of (\ref{eq.17}) and (\ref{eq.20}). In
order to obtain the number of  $2$-monomials, we have first to
determine the number of switch-off couples  $\alpha$ on $IS$. To do
it, depending of the parity of $K=(k'-1)(p+q)+i$, we have to
substract from $\Lambda_{1}$ (given by (\ref{eq.11}) or
(\ref{eq.12})) the population $\tilde{\Lambda}$ of all the couples
present on $IS$. As the cumulative multiplicity of all the
switch-off couples is equal to $p+q$, $\tilde{\Lambda}$ is divisible
by $p+q$. More precisely, one calculates $\Lambda_{1}$ by replacing
in (\ref{eq.11}) or (\ref{eq.12}) $N$ by $k'(p+q)+2+i$; one obtains:
\begin{equation}\label{eq.21} \Lambda_{1} =
\frac{1}{4}(k'-1)^{2}(p+q)^{2}+\frac{1}{2}(k'-1)(p+q)(i+2)+R,
\end{equation} with \begin{equation}\label{eq.22}
R=\frac{i^{2}}{4}+i+1 \,\,\text{($K$ even)}, \end{equation}
\begin{equation}\label{eq.23} R=\frac{i^{2}}{4}+i+\frac{3}{4}
\,\,\text{($K$ odd)}. \end{equation} For the case  $A$, $K$ has the
same parity as $i$ thus $R$ is canceled by $\tilde{\Lambda}''$ in
$\alpha=\Lambda_{1}-\tilde{\Lambda}'-\tilde{\Lambda}''$. We obtain:
\begin{eqnarray}\label{eq.24} \alpha &=&
\frac{1}{4}(k'-1)(k'-2)(p+q)^{2} \notag
\\
&+&\frac{1}{2}(k'-1)(i+2-\epsilon)(p+q).
\end{eqnarray}
To eliminate the redundancies in the switch-off
couples in $IS$, it is enough to divide  $\alpha$ by $p+q$. The
number of couples, thus the number of $2$-monomials $\Delta_{1}$ in
a sum $S_{m,\,\ell}^{(2)}$ ($m=-1,\,0$, $\ell=1,\,2$) is obtained by
summing $\frac{\alpha}{p+q}$ and $\tilde{\alpha}_{1}$, the number of
couples present without multiplicity on  $IS$ (to do it, one
attributes artificially a cumulative multiplicity of 1 to each of
the couples of the  $k'-1$ different classes $C_{k'-1}$ to $C_{1}$,
each cumulative multiplicity appearing exactly $1$ time in each
class, and to add the $\tilde{q}_{2}$ couples the class $C_{k'}$),
that is:
\begin{equation*} \tilde{\alpha_{1}} =
(\frac{p+q}{2})(k'-1)+E(\frac{i}{2})+1. \end{equation*} From which
we deduce that: $\Delta_{1} =
k'[1+E(\frac{i}{2})+\frac{(k'-1)(p+q)}{4}]$. Theorem $1$ is thus
proved.

\subsubsection{Case $B$}
\begin{table}[h]
\begin{center}
\begin{tabular}[h]{|c|c|c|} \hline
class $C_{j}$ & $q_{2}$ & $\mu$\\
\hline
$C_{k'-1}$ & $E(\frac{i}{2})+2$ & $p+q-1$        \\
$C_{k'-1}$ & $E(\frac{i}{2})+3$ & $p+q-3$  \\
\vdots & \vdots & \vdots \\
$C_{k'-1}$ & $E(\frac{i}{2})+\frac{p+q}{2}+\frac{1}{2}$ &  $2$ \\
\hline
$C_{k'-2}$ &  $E(\frac{i}{2})+\frac{p+q}{2}+\frac{3}{2}$ &$p+q$ \\
$C_{k'-2}$ &  $E(\frac{i}{2})+\frac{p+q}{2}+\frac{5}{2}$ &$p+q-2$ \\
\vdots & \vdots & \vdots \\
$C_{k'-2}$ & $E(\frac{i}{2})+p+q+1$ & $1$ \\
\hline
\vdots & \vdots & \vdots \\
\hline
$C_{2}$ & $E(\frac{i}{2})+\frac{(k'-3)(p+q)}{2}+\frac{3}{2}$ & $p+q$ \\
$C_{2}$ & $E(\frac{i}{2})+\frac{(k'-3)(p+q)}{2}+\frac{5}{2}$ & $p+q-2$ \\
\vdots & \vdots & \vdots \\
$C_{2}$ & $E(\frac{i}{2})+\frac{(k'-2)(p+q)}{2}+1$ & $1$ \\
\hline
$C_{1}$ & $E(\frac{i}{2})+\frac{(k'-2)(p+q)}{2}+2$ & $p+q-1$ \\
$C_{1}$ & $E(\frac{i}{2})+\frac{(k'-2)(p+q)}{2}+3$ & $p+q-3$ \\
\vdots & \vdots & \vdots  \\
$C_{1}$ & $E(\frac{i}{2})+\frac{(k'-1)(p+q)}{2}+\frac{1}{2}$ & $2$ \\
\hline
\end{tabular}
\caption{Case $B$. Table giving the different classes of couples
with cumulative multiplicities for $k'$ and $i$
even.}\label{table.Appendix.2}
\end{center}
\end{table}

Compared with case $A$, for a given class of couples, the cumulative
multiplicities are either all even or all odd. We have to
distinguish the  "sub-cases" $i$ even and $i$ odd: \begin{itemize}
\item If $i$ is even, one counts $\frac{p+q+1}{2}$ odd cumulative
multiplicities $1$, $3$, $...$, $p+q$, which appear exactly one time
on each of  the $\frac{k'-2}{2}$ even classes $C_{k'-2}$, $...$,
$C_{2}$ whereas the odd classes $C_{k'-1}$, $...$, $C_{1}$,
$\frac{k'}{2}$ in number, contain the $\frac{p+q-1}{2}$ even
cumulative multiplicities : $2$ to $p+q-1$. We denote
${\tilde{\Lambda}'}_{odd}$ and ${\tilde{\Lambda}'}_{even}$ the
number of couples of, respectively, odd and even cumulative
multiplicity on $IS$, except the couples of the class $C_{k'}$ which
population is given by (\ref{eq.17}). We have:
\begin{eqnarray}\label{eq.26}
{\tilde{\Lambda}'}_{odd} &=&
\frac{(k'-2)}{2}\sum_{j=0}^{\frac{p+q-1}{2}}(2j+1), \notag \\ & = &
\frac{(k'-2)(p+q+1)^{2}}{8}. \end{eqnarray}
\begin{eqnarray}\label{eq.27}
{\tilde{\Lambda}'}_{even} &=&
\frac{k'}{2}\sum_{j=1}^{\frac{p+q-1}{2}}(2j), \notag \\ & = &
\frac{k'(p+q-1)(p+q+1)}{8}.
\end{eqnarray}
One deduces:
\begin{equation}\label{eq.28}
\tilde{\Lambda}'  =
\frac{(k'-1)(p+q)^{2}}{4}+\frac{(k'-2)(p+q)}{4}-\frac{1}{4},
\end{equation}
then the population $\tilde{\Lambda}$ of all the
classes, the population of $C_{k'}$ being still given by Eq.
(\ref{eq.18}). The determination of switch-off couples is made
analogously to the case $A$. For $i$ and $k'$ even,
$K=(k'-1)(p+q)+i$ is odd, $\Lambda_{1}$ is obtained with
(\ref{eq.21}) and (\ref{eq.23}), thus it gives:
\begin{equation}\label{eq.29} \alpha =
\frac{1}{4}(k'-1)(k'-2)(p+q)^{2}+\frac{1}{4}[3k'+2i(k'-1)-2](p+q),
\end{equation}
One checks that (\ref{eq.29}) is divisible by $p+q$, $R$ being
canceled by the contributions (\ref{eq.18}) and $-\frac{1}{4}$ of
(\ref{eq.28}). Eq. (\ref{eq.29}) divided by $p+q$ and
$\tilde{\alpha_{1}}=(\frac{p+q}{2})(k'-1)+E(\frac{i}{2})+\frac{1}{2}$
gives: $\Delta_{1} = k'[\frac{(k'-1)(p+q)+2i+3}{4}].$

\item If $i$ is odd, there is $\frac{p+q+1}{2}$ odd cumulative
multiplicities  $1$, $3$, $...$, $p+q$, which appear exactly one
time on each of the  $\frac{k'}{2}$ odd  classes $C_{k'-1}$, $...$,
$C_{1}$ whereas the $\frac{(k'-2)}{2}$ even classes $C_{k'-2}$,
$...$, $C_{2}$ contain the $\frac{p+q-1}{2}$ even cumulative
multiplicity : $2$, $...$, $p+q-1$. We have now:
\begin{eqnarray}\label{eq.30} {\tilde{\Lambda}'}_{odd} &=&
\frac{k'}{2}\sum_{j=0}^{\frac{p+q-1}{2}}(2j+1), \notag \\ & = &
\frac{k'(p+q+1)^{2}}{8}. \end{eqnarray}
\begin{eqnarray}\label{eq.31} {\tilde{\Lambda}'}_{even} &=&
\frac{(k'-2)}{2}\sum_{j=1}^{\frac{p+q-1}{2}}(2j), \notag \\ & = &
\frac{(k'-2)(p+q-1)(p+q+1)}{8}. \end{eqnarray} It gives:
\begin{equation}\label{eq.32} \tilde{\Lambda}'  =
\frac{(k'-1)(p+q)^{2}}{4}+\frac{k'(p+q)}{4}+\frac{1}{4},
\end{equation} $\Lambda_{1}$ is determined by (\ref{eq.21}) and
(\ref{eq.22}), $\tilde{\Lambda}''$ by (\ref{eq.19}), from which one
has $\alpha=\Lambda_{1}-\tilde{\Lambda}' -\tilde{\Lambda}''$:
\begin{equation}\label{eq.33} \alpha =
\frac{1}{4}(k'-1)(k'-2)(p+q)^{2}+\frac{1}{4}[3k'+2i(k'-1)-4](p+q),
\end{equation} then the sum of
$\tilde{\alpha_{1}}=(\frac{p+q}{2})(k'-1)+E(\frac{i}{2})+\frac{3}{2}$
and (\ref{eq.33}) divided by  $p+q$ gives: $\Delta_{1} =
k'[\frac{(k'-1)(p+q)+2i+3}{4}]$.
\end{itemize}
\begin{table}[h]
\begin{center} \begin{tabular}[h]{|c|c|c|} \hline class $C_{j}$ &
$q_{2}$ & $\mu$\\ \hline $C_{k'-1}$ & $E(\frac{i}{2})+2$ & $p+q$ \\
$C_{k'-1}$ & $E(\frac{i}{2})+3$ & $p+q-2$        \\ \vdots & \vdots
& \vdots \\ $C_{k'-1}$ & $E(\frac{i}{2})+\frac{p+q}{2}+\frac{3}{2}$
& $1$ \\ \hline $C_{k'-2}$ &
$E(\frac{i}{2})+\frac{p+q}{2}+\frac{5}{2}$ & $p+q-1$ \\ $C_{k'-2}$ &
$E(\frac{i}{2})+\frac{p+q}{2}+\frac{7}{2}$ & $p+q-3$ \\ \vdots &
\vdots & \vdots  \\ $C_{k'-2}$ & $E(\frac{i}{2})+p+q+1$ & $2$ \\
\hline \vdots & \vdots & \vdots  \\ \hline $C_{2}$ &
$E(\frac{i}{2})+\frac{(k'-3)(p+q)}{2}+\frac{5}{2}$ & $p+q-1$ \\
$C_{2}$ & $E(\frac{i}{2})+\frac{(k'-3)(p+q)}{2}+\frac{7}{2}$ &
$p+q-3$ \\ \vdots & \vdots & \vdots \\ $C_{2}$ &
$E(\frac{i}{2})+\frac{(k'-2)(p+q)}{2}+1$ & $2$ \\ \hline $C_{1}$ &
$E(\frac{i}{2})+\frac{(k'-2)(p+q)}{2}+2$ & $p+q$ \\ $C_{1}$ &
$E(\frac{i}{2})+\frac{(k'-2)(p+q)}{2}+3$ & $p+q-2$ \\ \vdots &
\vdots & \vdots  \\ $C_{1}$ &
$E(\frac{i}{2})+\frac{(k'-1)(p+q)}{2}+\frac{3}{2}$ & $1$ \\ \hline
\end{tabular} \caption{Case $B$. Table giving the different classes
of couples with the cumulative multiplicities for $k'$ even and $i$
odd.}\label{table.Appendix.3} \end{center} \end{table} This proves
the theorem $2$ .

\subsubsection{Case $C$} As for case $B$, the study implies to
consider all the "sub-cases" $i$ even and $i$ odd: \begin{table}[h]
\begin{center} \begin{tabular}[h]{|c|c|c|} \hline class $C_{j}$ &
$q_{2}$ & $\mu$\\ \hline $C_{k'-1}$ & $E(\frac{i}{2})+2$ & $p+q-1$
\\ $C_{k'-1}$ & $E(\frac{i}{2})+3$ & $p+q-3$  \\ \vdots & \vdots &
\vdots \\ $C_{k'-1}$ & $E(\frac{i}{2})+\frac{p+q}{2}+\frac{1}{2}$ &
$2$ \\ \hline $C_{k'-2}$ &
$E(\frac{i}{2})+\frac{p+q}{2}+\frac{3}{2}$ &$p+q$ \\ $C_{k'-2}$ &
$E(\frac{i}{2})+\frac{p+q}{2}+\frac{5}{2}$ &$p+q-2$ \\ \vdots &
\vdots & \vdots \\ $C_{k'-2}$ & $E(\frac{i}{2})+p+q+1$ & $1$ \\
\hline \vdots & \vdots & \vdots \\ \hline $C_{2}$ &
$E(\frac{i}{2})+\frac{(k'-3)(p+q)}{2}+2$ & $p+q-1$ \\ $C_{2}$ &
$E(\frac{i}{2})+\frac{(k'-3)(p+q)}{2}+3$ & $p+q-3$ \\ \vdots &
\vdots & \vdots \\ $C_{2}$ &
$E(\frac{i}{2})+\frac{(k'-2)(p+q)}{2}+\frac{1}{2}$ & $2$ \\ \hline
$C_{1}$ & $E(\frac{i}{2})+\frac{(k'-2)(p+q)}{2}+\frac{3}{2}$ & $p+q$
\\ $C_{1}$ & $E(\frac{i}{2})+\frac{(k'-2)(p+q)}{2}+\frac{5}{2}$ &
$p+q-2$ \\ \vdots & \vdots & \vdots  \\ $C_{1}$ &
$E(\frac{i}{2})+\frac{(k'-1)(p+q)}{2}+1$ & $1$ \\ \hline
\end{tabular} \caption{Case $C$. Table giving the different classes
of couples with the cumulative multiplicity for $k'$ odd and $i$
even.}\label{table.Appendix.4} \end{center} \end{table}
\begin{itemize}
\item If $i$ is even, there is $\frac{p+q+1}{2}$ odd
cumulative multiplicities  $1$, $3$, $...$, $p+q$, which appear
exactly one time on each of the  $\frac{k'-1}{2}$ odd classes
$C_{k'-2}$, $...$, $C_{1}$, the $\frac{k'-1}{2}$ even classes
$C_{k'-1}$, $...$, $C_{2}$ having the $\frac{p+q-1}{2}$ even
cumulative multiplicities: $2$, $...$, $p+q-1$. It gives:
\begin{eqnarray}\label{eq.34}
{\tilde{\Lambda}'}_{odd} &=&
\frac{(k'-1)}{2}\sum_{j=0}^{\frac{p+q-1}{2}}(2j+1), \notag \\ & = &
\frac{(k'-1)(p+q+1)^{2}}{8}.
\end{eqnarray}
\begin{eqnarray}\label{eq.35}
{\tilde{\Lambda}'}_{even} &=&
\frac{(k'-1)}{2}\sum_{j=1}^{\frac{p+q-1}{2}}(2j), \notag \\ & = &
\frac{(k'-1)(p+q-1)(p+q+1)}{8}.
\end{eqnarray}

>From which we deduce:
\begin{equation}\label{eq.36}
\tilde{\Lambda}'  = \frac{(k'-1)(p+q)(p+q+1)}{4},
\end{equation}
then the population $\tilde{\Lambda}$ of all the classes, the population of
$C_{k'}$ is determined by (\ref{eq.18}). The number of switch-off couples is obtained on a analogous way that cases $A$ and $B$.
$K=(k'-1)(p+q)+i$ being even, $\Lambda_{1}$ is given by
(\ref{eq.21}) and (\ref{eq.22}), $\tilde{\Lambda}''$ is given by
(\ref{eq.18}), then
$\alpha=\Lambda_{1}-\tilde{\Lambda}'-\tilde{\Lambda}''$:
\begin{equation}\label{eq.37}
\alpha = \frac{1}{4}(k'-1)(p+q)[(k'-2)(p+q)+2i+3].
\end{equation}
Using Eq. (\ref{eq.37}) and $\tilde{\alpha_{1}} =
(\frac{p+q}{2})(k'-1)+E(\frac{i}{2})+1$, one deduces: $ \Delta_{1}
= (k'-1)[\frac{k'(p+q)+2i+3}{4}]+E(\frac{i}{2})+1$.

\item If $i$ is odd, there is $\frac{p+q+1}{2}$ odd multiplicities
$1$, $3$, $...$, $p+q$, which appear exactly one time on each of the
$\frac{(k'-1)}{2}$ even classes  $C_{k'-1}$, $...$, $C_{2}$ while
the  $\frac{(k'-1)}{2}$  odd classes $C_{k'-2}$ to $C_{1}$ contain
the $\frac{p+q-1}{2}$ even multiplicities $2$ to $p+q-1$.
Populations ${\tilde{\Lambda}'}_{odd}$, ${\tilde{\Lambda}'}_{even}$
and
$\tilde{\Lambda}'={\tilde{\Lambda}'}_{even}+{\tilde{\Lambda}'}_{odd}$
are still given by Eqs. (\ref{eq.34}), (\ref{eq.35}) and
(\ref{eq.36}); $\tilde{\Lambda}''$ is calculated by Eq.
(\ref{eq.19}) and $\Lambda_{1}$ deduced by Eqs. (\ref{eq.21}) and
(\ref{eq.23}). One obtains once again Eq. (\ref{eq.37}) for
$\alpha$. From Eq. (\ref{eq.37}) and
$\tilde{\alpha_{1}}=(\frac{p+q}{2})(k'-1)+E(\frac{i}{2})+1$, one may
write: \\
$\Delta_{1} =
(k'-1)[\frac{k'(p+q)+2i+3}{4}]+E(\frac{i}{2})+1$.
\end{itemize}

\begin{table}[h] \begin{center} \begin{tabular}[h]{|c|c|c|} \hline
Class $C_{j}$ & $q_{2}$ & $\mu$\\ \hline $C_{k'-1}$ &
$E(\frac{i}{2})+2$ & $p+q$        \\ $C_{k'-1}$ & $E(\frac{i}{2})+3$
& $p+q-2$        \\ \vdots & \vdots & \vdots \\ $C_{k'-1}$ &
$E(\frac{i}{2})+\frac{p+q}{2}+\frac{3}{2}$ & $1$ \\ \hline
$C_{k'-2}$ &  $E(\frac{i}{2})+\frac{p+q}{2}+\frac{5}{2}$ & $p+q-1$
\\ $C_{k'-2}$ &  $E(\frac{i}{2})+\frac{p+q}{2}+\frac{7}{2}$ &
$p+q-3$ \\ \vdots & \vdots & \vdots  \\ $C_{k'-2}$ &
$E(\frac{i}{2})+p+q+1$ & $2$ \\ \hline \vdots & \vdots & \vdots  \\
\hline $C_{2}$ & $E(\frac{i}{2})+\frac{(k'-3)(p+q)}{2}+2$ & $p+q$ \\
$C_{2}$ & $E(\frac{i}{2})+\frac{(k'-3)(p+q)}{2}+3$ & $p+q-2$ \\
\vdots & \vdots & \vdots \\ $C_{2}$ &
$E(\frac{i}{2})+\frac{(k'-2)(p+q)}{2}+\frac{3}{2}$ & $1$ \\ \hline
$C_{1}$ & $E(\frac{i}{2})+\frac{(k'-2)(p+q)}{2}+\frac{5}{2}$ &
$p+q-1$ \\ $C_{1}$ &
$E(\frac{i}{2})+\frac{(k'-2)(p+q)}{2}+\frac{7}{2}$ & $p+q-3$ \\
\vdots & \vdots & \vdots  \\ $C_{1}$ &
$E(\frac{i}{2})+\frac{(k'-1)(p+q)}{2}+1$ & $2$ \\ \hline
\end{tabular} \caption{Case $C$. Table giving the different classes
of couples with the cumulative multiplicities for $k'$ and $i$
odd.}\label{table.Appendix.5} \end{center} \end{table} Theorem $3$
is thus demonstrated.

\subsection{$3$-monomials counting}
In this section, we give the demonstration of the three
$3$-monomials counting theorems. For the following, without limiting
the generality of the problem, the order $N$ of the expansion
(\ref{eq.100}) is denoted $N=k'(p+q)+4+i$ with $k'$ and
$i\in[0,\,...,\,p+q-1]$ two strictly positive integers. We work on
the secondary interval $IS=[2,\,Q_{3}]_{N=k'(p+q)+4+i}$ with
$Q_{3}=E(\frac{(k'-1)(p+q)+4+i}{2})$.

\subsubsection{The different classes of couples}
Successive values
taken by the integer $q_{3}$ on $IS$ compose the classes of the
couples : $C_{1},\,...,\,C_{k'-2},\,\,C_{k'-1}$ and $C_{k'}$
(classes $C_{k'-1}$ and $C_{k'-2}$ exist if, respectively,  $k'\geq
2$ and $k'\geq 3$). We begin by giving explicitly the $3$-couples
belonging to the class $C_{k'}$ as well as its population
$\tilde{\Lambda}''$. The $3$-couples $(k',\,q_{3},\,\gamma)$
($\gamma=1,\,...,\,q_{3}-1$) of this class are: $(k',\,1,\,\gamma)$,
$(k',\,2,\,\gamma)$, $...$, $(k',\tilde{q}_{3},\,\gamma)$ with
$\tilde{q}_{3}$ the highest integer value of $q_{3}$ allowed such
that $p_{3} = E(\frac{k'(p+q)+4+i-2\tilde{q}_{3}}{p+q})=k'$, that is
$\tilde{q}_{3}=2+E(\frac{i}{2})$. We determine the  multiplicities
of these $3$-couples with the help of (\ref{eq.15}) and
(\ref{eq.16}). Thus, the cumulative multiplicities of the couples
$(k',\,1,\,\gamma)$, $(k',\,2,\,\gamma)$, $...$,
$(k',\tilde{q}_{3},\,\gamma)$ on $IS$ are, respectively,
$\mu=i+1,\,i-1,\,...,\,\,i+5-2q_{3},\,...,\,1+\epsilon$ with
$\epsilon=i-2E(\frac{i}{2})$. However, to a given value of $q_{3}$,
there is $q_{3}-1$ $3$-couples $(k',\,q_{3},\,\gamma)$. The counting
of the population of  $C_{k'}$ is equivalent to the counting  of all
the $2$-couples $(k',\,q_{3})$  which cumulative multiplicities on
$IS$ is the product of one of the  $q_{3}-1$ $3$-couples
$(k',\,q_{3},\,\gamma)$ which it relates and of $q_{3}-1$. It reads:
\begin{eqnarray}\label{eq.41} \tilde{\Lambda}'' &=&
\sum_{q_{3}=2}^{\tilde{q}_{3}}(i+5-2q_{3})(q_{3}-1), \notag \\ &=&
\frac{1}{2}(i+7)[E(\frac{i}{2})+1][E(\frac{i}{2})+4]+2-(i+5)[E(\frac{i}{2})+1]\notag
\\
&-&\frac{1}{3}[E(\frac{i}{2})+2][E(\frac{i}{2})+3][2E(\frac{i}{2})+5],
\end{eqnarray} or also, depending of the parity of $i$:
\begin{equation}\label{eq.42}
\tilde{\Lambda}''=\frac{i^{3}}{24}+\frac{3i^{2}}{8}+\frac{13i}{12}+1\,\,\text{($i$
even)}, \end{equation} \begin{equation}\label{eq.43}
\tilde{\Lambda}''=\frac{i^{3}}{24}+\frac{3i^{2}}{8}+\frac{23i}{24}+\frac{5}{8}\,\,\text{($i$
odd)}. \end{equation} The $3$-couples of the others different
classes of $IS$ divide as follows: $(k'-1,
E(\frac{i}{2})+3,\,\gamma)$, $(k'-1, E(\frac{i}{2})+4,\,\gamma)$,
$...$, $(k'-1, E(\frac{p+q+i+4}{2}),\,\gamma)$ for the class
$C_{k'-1}$, $(k'-2, E(\frac{p+q+i+6}{2}),\,\gamma)$, $(k'-2,
E(\frac{p+q+i+8}{2}),\,\gamma)$, $...$,  $(k'-2, \\
E(\frac{2(p+q)+i+4}{2}),\,\gamma)$ for the class $C_{k'-2}$, $...$,
$(2, E(\frac{(k'-3) (p+q)+i+6}{2}) \\ ,\,\gamma)$, $(2,
E(\frac{(k'-3)(p+q)+i+ 8}{2}),\,\gamma)$, $...$, $(2,
E(\frac{(k'-2)(p+q) +i+4}{2}),\,\gamma)$ for the class $C_{2}$ and
finally $(1, E(\frac{(k'-2) (p+q)+i+6}{2}),\,\gamma)$,  $(1, \\
E(\frac{(k'-2)(p+q) +i+ 8}{2}),\,\gamma)$, $...$, $(1,
Q_{3}=E(\frac{(k'-1) (p+q)+i+4}{2}),\,\gamma)$ for the class
$C_{1}$. As for the $2$-monomials, the calculation of a cumulative
multiplicity implies to distinguish the three cases: $p+q$ even
(case A), $p+q$ odd with  $k'$ even (case B) and $p+q$ odd with $k'$
odd (case C). Results are given in the tables \ref{table.Appendix.6}
to \ref{table.Appendix.10}.

\begin{table}[h] \begin{center} \begin{tabular}[h]{|c|c|c|} \hline
class $C_{j}$ & $q_{3}$  &  $\mu$ \\ \hline $C_{k'-1}$ &
$E(\frac{i}{2})+3$ &  $p+q-1+\epsilon$         \\ $C_{k'-1}$ &
$E(\frac{i}{2})+4$ & $p+q-3+\epsilon$         \\ \vdots & \vdots &
\vdots \\ $C_{k'-1}$ & $E(\frac{i}{2})+\frac{p+q}{2}+2$ &
$1+\epsilon$ \\ \hline $C_{k'-2}$ &$E(\frac{i}{2})+\frac{p+q}{2}+3$
& $p+q-1+\epsilon$ \\ $C_{k'-2}$ &$E(\frac{i}{2})+\frac{p+q}{2}+4$ &
$p+q-3+\epsilon$ \\ \vdots & \vdots & \vdots \\ $C_{k'-2}$ &
$E(\frac{i}{2})+p+q+2$ & $1+\epsilon$ \\ \hline \vdots & \vdots &
\vdots \\ \hline $C_{2}$ & $E(\frac{i}{2})+\frac{(k'-3)(p+q)}{2}+3$
& $p+q-1+\epsilon$ \\ $C_{2}$ &
$E(\frac{i}{2})+\frac{(k'-3)(p+q)}{2}+4$ & $p+q-3+\epsilon$ \\
\vdots & \vdots & \vdots \\ $C_{2}$ &
$E(\frac{i}{2})+\frac{(k'-2)(p+q)}{2}+2$ & $1+\epsilon$ \\ \hline
$C_{1}$ & $E(\frac{i}{2})+\frac{(k'-2)(p+q)}{2}+3$ &
$p+q-1+\epsilon$ \\ $C_{1}$ &
$E(\frac{i}{2})+\frac{(k'-2)(p+q)}{2}+4$ & $p+q-3+\epsilon$ \\
\vdots & \vdots & \vdots \\ $C_{1}$ &
$E(\frac{i}{2})+\frac{(k'-1)(p+q)}{2}+2$ & $1+\epsilon$ \\ \hline
\end{tabular} \caption{Case $A$. Table giving the different classes
of $3$-couples as a function of  $q_{3}$ with the cumulative
multiplicities. If $i$ even ($\epsilon=0$), all the cumulative
multiplicities are odd; they are all even if $i$ is odd
($\epsilon=1$).}\label{table.Appendix.6} \end{center} \end{table}

\subsubsection{Case $A$} We denote by $\tilde{\Lambda}'$ the
population of the classes $C_{k'-1}$ to $C_{1}$; the counting of
$\tilde{\Lambda}'$ is more tedious than for the $2$-monomials,
because, for one given class of couples, there is $q_{3}-1$
$3$-couples of same multiplicity on $IS$ and this value varies from
a class of couples to another. The method used here, consists in
counting successively  the populations of the $\frac{p+q}{2}$
classes of multiplicity $\tilde{\Lambda}_{p+q-1+\epsilon}$,
$\tilde{\Lambda}_{p+q-3+\epsilon}$, $...$,
$\tilde{\Lambda}_{1+\epsilon}$. For instance, there is
$E(\frac{i}{2})+1+\frac{j(p+q)}{2}$ $3$-couples ($1\leq j \leq
k'-1$) of cumulative multiplicity $\mu=1+\epsilon$ in each of the
classes  $C_{j}$. Thus one has:
\begin{eqnarray}\label{eq.44}
\tilde{\Lambda}_{1+\epsilon} &=&
(1+\epsilon)\sum_{j=1}^{k'-1}[E(\frac{i}{2})+1+\frac{j(p+q)}{2}],
\notag \\ & = &
(1+\epsilon)(k'-1)[E(\frac{i}{2})+1+\frac{k'(p+q)}{4}].
\end{eqnarray}
Doing similarly for the others populations:
\begin{eqnarray} \tilde{\Lambda}_{3+\epsilon} &=&
(3+\epsilon)\sum_{j=1}^{k'-1}[E(\frac{i}{2})+\frac{j(p+q)}{2}],
\notag \\ & = &
(3+\epsilon)(k'-1)[E(\frac{i}{2})+\frac{k'(p+q)}{4}], \label{eq.45}
\\ &\vdots& \notag \\ \tilde{\Lambda}_{p+q-1+\epsilon} &=&
(p+q-1+\epsilon)\sum_{j=0}^{k'-2}[E(\frac{i}{2})+2+\frac{j(p+q)}{2}],
\notag \\ & = & (p+q-1+\epsilon)(k'-1)[E(\frac{i}{2})+2 \notag \\
&+&\frac{(k'-2)(p+q)}{4}]. \label{eq.46}
\end{eqnarray}
Population
$\tilde{\Lambda}'$ is the sum of Eqs. (\ref{eq.44}) to
(\ref{eq.46}):
\begin{eqnarray}\label{eq.47} \tilde{\Lambda}' &=&
\sum_{j=0}^{\frac{p+q-2}{2}}(\epsilon+2j+1)(k'-1)[E(\frac{i}{2})
\notag \\
&+& 1-j+\frac{k'(p+q)}{4}], \notag \\ & = &
\frac{(k'-1)(p+q)}{2}[(\epsilon+\frac{p+q}{2})(E(\frac{i}{2})+\frac{k'(p+q)}{4})
\notag \\ &+&\frac{(p+q-2)}{12}(5-2\epsilon-2(p+q))+\epsilon+1].
\end{eqnarray}
The population $\tilde{\Lambda}=\tilde{\Lambda}'+\tilde{\Lambda}''$
of all the classes $C_{j}$ ($j=1,\,...,\,k'$) is the sum of Eqs.
(\ref{eq.41}) and (\ref{eq.47}). To count the number of
$3$-monomials in a sum $S_{m}^{(3)}$, one has at first to determine
the number of switch-off couples  $\alpha$ on $IS$, the method
remaining the same as for the $2$-monomials. It gives:
$\alpha=\Lambda_{2}-\tilde{\Lambda}$, where depending on the parity
of $K=(k'-1)(p+q)+i$, $\Lambda_{2}$ is given by Eq. (\ref{eq.13}) or
(\ref{eq.14}). In Eqs. (\ref{eq.13})  or (\ref{eq.14}) we replace
$N$ by $k'(p+q)+4+i$, and one obtains the following equations:
\begin{eqnarray}\label{eq.48} \Lambda_{2} &=&
\frac{1}{24}(k'-1)^{3}(p+q)^{3}+\frac{1}{8}(k'-1)^{2}(p+q)^{2}(i+3)\notag
\\ &+& \frac{1}{24}(k'-1)(p+q)(3i(i+6)+26-3\epsilon)\notag \\ &+&R,
\end{eqnarray} \begin{equation}\label{eq.49}
R=\frac{i^{3}}{24}+\frac{3i^{2}}{8}+\frac{13i}{12}+1\,\,\text{($K$
even)}, \end{equation} \begin{equation}\label{eq.50}
R=\frac{i^{3}}{24}+\frac{3i^{2}}{8}+\frac{23i}{24}+\frac{5}{8}\,\,\text{($K$
odd)}. \end{equation} For the case  $A$, $K$  and $i$ have the same
parity, thus  $R$ is canceled by $\tilde{\Lambda}''$. All
calculations made, it results: \begin{eqnarray}\label{eq.51} \alpha
&=&
\frac{1}{24}(k'-1)(p+q)\biggl(\frac{1}{2}(p+q)^{2}[2(k'-1)^{2}-3k']
\notag \\ &+& (p+q)[3(k'-1)(i+3)-3\epsilon k'+2(p+q-2) \notag \\ &-&
6E(\frac{i}{2})]+2(3+i)^{2}+(4+i)(2+i) \notag \\ &-& 12\epsilon
E(\frac{i}{2}) -(p+q-2)(5-\epsilon)-12(\epsilon+1)\biggr).
\end{eqnarray} The number of $3$-monomials $\Delta_{2}$ in a sum
$S_{m}^{(3)}$, is obtained by summing $\frac{\alpha}{p+q}$ and
$\tilde{\alpha}_{2}$, the number of couples present without
multiplicity on $IS$ take into account the $3$-couples of the
classes $C_{j}$ ($1\leq j \leq k'-1$) and of the class $C_{k'}$:
\begin{eqnarray*} \tilde{\alpha_{2}} &=&
\sum_{j=0}^{\frac{p+q-2}{2}}[E(\frac{i}{2})+1+\frac{k'(p+q)}{4}-j]+\sum_{q_{3}=2}^{\tilde{q}_{3}}(q_{3}-1),
\notag \\ &=&
\frac{[E(\frac{i}{2})+1][E(\frac{i}{2})+2]}{2}+\frac{(k'-1)(p+q)}{2}[E(\frac{i}{2})
\notag \\ &+&\frac{3}{2}+\frac{(k'-1)(p+q)}{4}] \end{eqnarray*} One
deduces  $\Delta_{2} = \frac{\alpha}{p+q}+\tilde{\alpha_{2}}$ given
by Eq. (\ref{theor.3a}); this proves theorem $4$.

\subsubsection{Case $B$}
\begin{table}[h]
\begin{center}
\begin{tabular}[h]{|c|c|c|} \hline
class $C_{j}$ & $q_{3}$ & $\mu$\\
\hline
$C_{k'-1}$ & $E(\frac{i}{2})+3$ & $p+q-1$        \\
$C_{k'-1}$ & $E(\frac{i}{2})+4$ & $p+q-3$  \\
\vdots & \vdots & \vdots \\
$C_{k'-1}$ & $E(\frac{i}{2})+\frac{p+q}{2}+\frac{3}{2}$ &  $2$ \\
\hline
$C_{k'-2}$ &  $E(\frac{i}{2})+\frac{p+q}{2}+\frac{5}{2}$ &$p+q$ \\
$C_{k'-2}$ &  $E(\frac{i}{2})+\frac{p+q}{2}+\frac{7}{2}$ &$p+q-2$ \\
\vdots & \vdots & \vdots \\
$C_{k'-2}$ & $E(\frac{i}{2})+p+q+2$ & $1$ \\
\hline
\vdots & \vdots & \vdots \\
\hline
$C_{2}$ & $E(\frac{i}{2})+\frac{(k'-3)(p+q)}{2}+\frac{5}{2}$ & $p+q$ \\
$C_{2}$ & $E(\frac{i}{2})+\frac{(k'-3)(p+q)}{2}+\frac{7}{2}$ & $p+q-2$ \\
\vdots & \vdots & \vdots \\
$C_{2}$ & $E(\frac{i}{2})+\frac{(k'-2)(p+q)}{2}+2$ & $1$ \\
\hline
$C_{1}$ & $E(\frac{i}{2})+\frac{(k'-2)(p+q)}{2}+3$ & $p+q-1$ \\
$C_{1}$ & $E(\frac{i}{2})+\frac{(k'-2)(p+q)}{2}+4$ & $p+q-3$ \\
\vdots & \vdots & \vdots  \\
$C_{1}$ & $E(\frac{i}{2})+\frac{(k'-1)(p+q)}{2}+\frac{3}{2}$ & $2$ \\
\hline
\end{tabular}
\caption{Case $B$. Table giving the different classes of
$3$-couples function of  $q_{3}$ with the cumulative multiplicities
for $k'$ and $i$ even.}\label{table.Appendix.7}
\end{center}
\end{table}

Compared with case $A$, as for the $2$-monomials, one has to
consider the "sub-cases" $i$ even and $i$ odd. Once again, the
method consists to begin with the populations of the different
classes of multiplicity, to determine ${\tilde{\Lambda}'}_{even}$
and ${\tilde{\Lambda}'}_{odd}$, the populations of even and odd
cumulative multiplicities on $IS$, except the class $C_{k'}$ which
population is known and given by Eq. (\ref{eq.41}). \begin{itemize}
\item If $i$ is even, there is $\frac{p+q+1}{2}$ classes of odd
multiplicities with populations $\tilde{\Lambda}_{2j+1}$ ($0\leq
j\leq \frac{p+q-1}{2}$), each of these classes appears 1 time on
each of the $\frac{k'-2}{2}$ even classes of couples $C_{k'-2}$,
$...$, $C_{2}$. On the set of these classes, one counts
$E(\frac{i}{2})+\frac{k'(p+q)}{4}+1-j$ $3$-couples of cumulative
multiplicity $\mu=2j+1$. It gives: \begin{eqnarray}\label{eq.53}
{\tilde{\Lambda}'}_{odd} &=&
\frac{(k'-2)}{2}\sum_{j=0}^{\frac{p+q-1}{2}}(2j+1)(E(\frac{i}{2})
\notag \\ &+&\frac{k'(p+q)}{4}+1-j) \notag \\ &=&
\frac{(k'-2)(p+q+1)}{4}\biggl(E(\frac{i}{2})+\frac{k'(p+q)}{4}
\notag \\
&+&1+\frac{(p+q-1)}{4}[2E(\frac{i}{2})+1+\frac{k'(p+q)}{2}] \notag
\\ &-&\frac{1}{6}(p+q)(p+q-1)\biggr). \end{eqnarray} One counts also
$\frac{p+q-1}{2}$ classes of even multiplicity with populations
$\tilde{\Lambda}_{2j}$ ($1\leq j\leq \frac{p+q-1}{2}$), each of
these classes appearing 1 time on each of the $\frac{k'}{2}$ odd
classes of couples $C_{k'-1}$, $...$, $C_{1}$. Furthermore , to a
given cumulative multiplicity $\mu=2j$, correspond
$E(\frac{i}{2})+\frac{k'(p+q)}{4}+\frac{3}{2}-j$ $3$-couples, thus:
\begin{eqnarray}\label{eq.54} {\tilde{\Lambda}'}_{even} &=&
\frac{k'}{2}\sum_{j=1}^{\frac{p+q-1}{2}}(2j)(E(\frac{i}{2}) +
\frac{k'(p+q)}{4}+\frac{3}{2}-j) \notag \\ &=&
\frac{k'(p+q+1)}{4}\biggl(\frac{(p+q-1)}{4}[2E(\frac{i}{2})+3 \notag
\\ &+&\frac{k'(p+q)}{2}] - \frac{1}{6}(p+q)(p+q-1)\biggr).
\end{eqnarray} One deduces the population of all the classes $C_{j}$
($j=1,\,...,\,k'-1$): \begin{eqnarray}\label{eq.55} \tilde{\Lambda}'
&=& \frac{(k'-1)(3k'-4)(p+q)^{3}}{48}\notag \\
&+&\frac{(k'-1)(p+q)^{2}}{8}[2E(\frac{i}{2})+1]+\frac{k'^{2}(p+q)^{2}}{16}\notag
\\
&+&\frac{(p+q)}{4}[(k'-2)E(\frac{i}{2})+\frac{13k'}{12}-\frac{7}{3}]
\notag \\ &-&\frac{1}{4}E(\frac{i}{2})-\frac{3}{8}. \end{eqnarray}
For $i$ and $k'$ even, $K=(k'-1)(p+q)+i$ is odd, $\Lambda_{2}$ is
obtained by Eqs. (\ref{eq.48}) and (\ref{eq.50}), thus it results:
\begin{eqnarray}\label{eq.56} \alpha &=&
\frac{(k'-1)(k'-2)(2k'-3)(p+q)^{3}}{48}\notag \\
&+&\frac{(k'-1)(p+q)^{2}}{8}[(k'-1)(3+i)-2E(\frac{i}{2})-1]\notag \\
&-&\frac{k'^{2}(p+q)^{2}}{16}
+\frac{(p+q)}{4}\biggl(\frac{(k'-1)}{6}[3i(i+6)+23] \notag \\
&-&(k'-2)E(\frac{i}{2}) -\frac{13k'}{12}+\frac{7}{3}\biggr).
\end{eqnarray}
Eq.(\ref{eq.56}) is divisible by $p+q$, the sum of terms
$-\frac{1}{4}E(\frac{i}{2})-\frac{3}{8}$ in Eq. (\ref{eq.55}) and of
$\tilde{\Lambda}''$ (Eq. (\ref{eq.42})) cancels $R$
(Eq.(\ref{eq.50})). Moreover, the number of $3$-couples present
without multiplicity on $IS$ is: \begin{eqnarray*}
\tilde{\alpha_{2}}&=&
\frac{[E(\frac{i}{2})+1][E(\frac{i}{2})+2]}{2}\notag \\ &+&
\frac{(k'-2)}{2}\sum_{j=0}^{\frac{p+q-1}{2}}[E(\frac{i}{2})+\frac{k'(p+q)}{4}+1-j]
\notag \\
&+&\frac{k'}{2}\sum_{j=1}^{\frac{p+q-1}{2}}[E(\frac{i}{2})+\frac{k'(p+q)}{4}+\frac{3}{2}-j]
\notag \\ &=& \frac{[E(\frac{i}{2})+1][E(\frac{i}{2})+2]}{2} \notag
\\ &+&
\frac{(k'-1)(p+q)}{2}[\frac{(k'-1)(p+q)}{4}+E(\frac{i}{2})+1]\notag
\\ &-&\frac{E(\frac{i}{2})}{2}-\frac{5}{8}. \end{eqnarray*} We
obtain: \begin{eqnarray}\label{eq.58} \Delta_{2} &=&
\frac{[E(\frac{i}{2})+1][E(\frac{i}{2})+2]}{2} \notag \\ &+&
\frac{k'(k'-1)(p+q)}{48}[(2k'-1)(p+q)+3(2i+5)] \notag \\ &+&
\frac{(k'-1)}{8}[i(i+6)+8] -\frac{k'}{16}[4E(\frac{i}{2}) \notag \\
&+& (p+q)+5]. \end{eqnarray} This is Eq.(\ref{theor.3b}) for $i$
even ($\epsilon=0$). \item If $i$ is odd, one counts
$\frac{p+q+1}{2}$ classes of odd multiplicity with populations
$\tilde{\Lambda}_{2j+1}$ ($0\leq j\leq \frac{p+q-1}{2}$), each of
these classes appearing one time on each of the $\frac{k'}{2}$ odd
classes of couples $C_{k'-1}$, $...$, $C_{1}$. On the set of these
classes, one counts $E(\frac{i}{2})+\frac{k'(p+q)}{4}+\frac{3}{2}-j$
$3$-couples of cumulative multiplicity $\mu=2j+1$. It reads:
\begin{eqnarray}\label{eq.59} {\tilde{\Lambda}'}_{odd} &=&
\frac{k'(p+q+1)}{24}\biggl(3\epsilon[2E(\frac{i}{2})+3]+\frac{3\epsilon
k'(p+q)}{2} \notag \\
&-&(p+q-1)(p+q)+\frac{3(p+q-1)}{2}[2E(\frac{i}{2}) \notag \\
&+&\frac{k'(p+q)}{2}+3-\epsilon]\biggr). \end{eqnarray} There is
also $\frac{p+q-1}{2}$ classes of even multiplicity with
 populations $\tilde{\Lambda}_{2j}$ ($0\leq j\leq \frac{p+q-1}{2}$),
each of these classes appearing one time on each of the
$\frac{(k'-2)}{2}$ even classes of couples $C_{k'-2}$, $...$,
$C_{2}$. Furthermore, for a given cumulative multiplicity $\mu=2j$,
there is $E(\frac{i}{2})+\frac{k'(p+q)}{4}+2-j$ $3$-couples. It
gives: \begin{eqnarray}\label{eq.60} {\tilde{\Lambda}'}_{even} &=&
\frac{(k'-2)(p+q-1)}{4}\biggl(\frac{(p+q-3)}{4}[2E(\frac{i}{2})+
\notag \\ &&\frac{k'(p+q)}{2}+1-\epsilon]
+(1+\epsilon)[E(\frac{i}{2})+1 \notag \\ &+&
\frac{k'(p+q)}{4}]-\frac{(p+q-2)(p+q-3)}{6}\biggr). \end{eqnarray}
One deduces successively the populations $\tilde{\Lambda}$ of all
the classes $C_{j}$ ($j=1,\,...,\,k'$) by summing Eqs.
(\ref{eq.43}), (\ref{eq.59}) and (\ref{eq.60}), then $\Lambda_{2}$
with the help of Eqs. (\ref{eq.48}) and (\ref{eq.49}), $\alpha =
\Lambda_{2}-\tilde{\Lambda}$. One determines then the number of
$3$-couples present without  multiplicity on $IS$ by:
\begin{eqnarray}\label{eq.61} \tilde{\alpha_{2}}&=&
\frac{[E(\frac{i}{2})+1][E(\frac{i}{2})+2]}{2} \notag \\
&+&\frac{(k'-1)(p+q)}{2}[\frac{(k'-1)(p+q)}{4}+E(\frac{i}{2})+2]\notag
\\ &+&\frac{E(\frac{i}{2})}{2}+\frac{7}{8}, \end{eqnarray} It
follows that: \begin{eqnarray}\label{eq.62} \Delta_{2} &=&
\frac{[E(\frac{i}{2})+1][E(\frac{i}{2})+2]}{2} \notag \\ &+&
\frac{k'(k'-1)(p+q)}{48}[(2k'-1)(p+q)+3(2i+5)] \notag \\ &+&
\frac{(k'-1)}{8}[i(i+6)-4 E(\frac{i}{2})+1] +
\frac{k'}{16}[4E(\frac{i}{2}) \notag \\ &+&(p+q)+7]. \end{eqnarray}
$\Delta_{2}$ is deduced from (\ref{theor.3b}) for the value
$\epsilon=1$. Theorem $5$ is demonstrated.
\end{itemize}
\begin{table}[h] \begin{center} \begin{tabular}[h]{|c|c|c|} \hline
class $C_{j}$ & $q_{3}$ & $\mu$\\ \hline $C_{k'-1}$ &
$E(\frac{i}{2})+3$ & $p+q$        \\ $C_{k'-1}$ & $E(\frac{i}{2})+4$
& $p+q-2$  \\ \vdots & \vdots & \vdots \\ $C_{k'-1}$ &
$E(\frac{i}{2})+\frac{p+q}{2}+\frac{5}{2}$ &  $1$ \\ \hline
$C_{k'-2}$ &  $E(\frac{i}{2})+\frac{p+q}{2}+\frac{7}{2}$ &$p+q-1$ \\
$C_{k'-2}$ &  $E(\frac{i}{2})+\frac{p+q}{2}+\frac{9}{2}$ &$p+q-3$ \\
\vdots & \vdots & \vdots \\ $C_{k'-2}$ & $E(\frac{i}{2})+p+q+2$ &
$2$ \\ \hline \vdots & \vdots & \vdots \\ \hline $C_{2}$ &
$E(\frac{i}{2})+\frac{(k'-3)(p+q)}{2}+\frac{7}{2}$ & $p+q-1$ \\
$C_{2}$ & $E(\frac{i}{2})+\frac{(k'-3)(p+q)}{2}+\frac{9}{2}$ &
$p+q-3$ \\ \vdots & \vdots & \vdots \\ $C_{2}$ &
$E(\frac{i}{2})+\frac{(k'-2)(p+q)}{2}+2$ & $2$ \\ \hline $C_{1}$ &
$E(\frac{i}{2})+\frac{(k'-2)(p+q)}{2}+3$ & $p+q$ \\ $C_{1}$ &
$E(\frac{i}{2})+\frac{(k'-2)(p+q)}{2}+4$ & $p+q-2$ \\ \vdots &
\vdots & \vdots  \\ $C_{1}$ &
$E(\frac{i}{2})+\frac{(k'-1)(p+q)}{2}+\frac{5}{2}$ & $1$ \\ \hline
\end{tabular} \caption{Case $B$. Table giving the  different classes
of $3$-couples as a function of $q_{3}$ with the cumulative
multiplicities for $k'$ even and $i$ odd
($\epsilon=1$).}\label{table.Appendix.8} \end{center} \end{table}

\subsubsection{Case $C$} \begin{table}[h] \begin{center}
\begin{tabular}[h]{|c|c|c|} \hline class $C_{j}$ & $q_{3}$ & $\mu$\\
\hline $C_{k'-1}$ & $E(\frac{i}{2})+3$ & $p+q-1$        \\
$C_{k'-1}$ & $E(\frac{i}{2})+4$ & $p+q-3$  \\ \vdots & \vdots &
\vdots \\ $C_{k'-1}$ & $E(\frac{i}{2})+\frac{p+q}{2}+\frac{3}{2}$ &
$2$ \\ \hline $C_{k'-2}$ &
$E(\frac{i}{2})+\frac{p+q}{2}+\frac{5}{2}$ &$p+q$ \\ $C_{k'-2}$ &
$E(\frac{i}{2})+\frac{p+q}{2}+\frac{7}{2}$ &$p+q-2$ \\ \vdots &
\vdots & \vdots \\ $C_{k'-2}$ & $E(\frac{i}{2})+p+q+2$ & $1$ \\
\hline \vdots & \vdots & \vdots \\ \hline $C_{2}$ &
$E(\frac{i}{2})+\frac{(k'-3)(p+q)}{2}+3$ & $p+q-1$ \\ $C_{2}$ &
$E(\frac{i}{2})+\frac{(k'-3)(p+q)}{2}+4$ & $p+q-3$ \\ \vdots &
\vdots & \vdots \\ $C_{2}$ &
$E(\frac{i}{2})+\frac{(k'-2)(p+q)}{2}+\frac{3}{2}$ & $2$ \\ \hline
$C_{1}$ & $E(\frac{i}{2})+\frac{(k'-2)(p+q)}{2}+\frac{5}{2}$ & $p+q$
\\ $C_{1}$ & $E(\frac{i}{2})+\frac{(k'-2)(p+q)}{2}+\frac{7}{2}$ &
$p+q-2$ \\ \vdots & \vdots & \vdots  \\ $C_{1}$ &
$E(\frac{i}{2})+\frac{(k'-1)(p+q)}{2}+2$ & $1$ \\ \hline
\end{tabular} \caption{Case $C$. Table giving the different classes
of $3$-couples as a function of $q_{3}$ with the cumulative
multiplicities for $k'$ odd and $i$ even.}\label{table.Appendix.9}
\end{center} \end{table} As for the case $B$, One has to consider
the "sub-cases" $i$ even and $i$ odd.

\begin{itemize} \item If $i$ is even, there is $\frac{p+q+1}{2}$
classes of odd multiplicity with the populations
$\tilde{\Lambda}_{2j+1}$ ($0\leq j\leq \frac{p+q-1}{2}$), each of
these classes appearing exactly one time on each of the
$\frac{k'-1}{2}$ odd classes of couples $C_{k'-2}$, $...$, $C_{1}$.
Furthermore, for a given cumulative multiplicity $\mu=2j+1$, there
is $E(\frac{i}{2})+\frac{(k'+1)(p+q)}{4}+1-j$ $3$-couples. Thus one
may write: \begin{eqnarray}\label{eq.63} {\tilde{\Lambda}'}_{odd}
&=& \frac{(k'-1)(p+q+1)}{4}\biggl(E(\frac{i}{2}) \notag \\
&+&\frac{(k'+1)(p+q)}{4}+1+\frac{(p+q-1)}{4}[2E(\frac{i}{2}) \notag
\\ &+&1+\frac{(k'+1)(p+q)}{2}] \notag \\ &-&
\frac{1}{6}(p+q)(p+q-1)\biggr).
\end{eqnarray}
There is also
$\frac{p+q-1}{2}$ classes of even multiplicity with populations
$\tilde{\Lambda}_{2j}$ ($1\leq j\leq \frac{p+q-1}{2}$), each of
these classes appearing exactly one time on each of the
$\frac{(k'-1)}{2}$ even classes of couples $C_{k'-1}$, $...$,
$C_{2}$. Furthermore, for a given cumulative multiplicity $\mu=2j$,
there is $E(\frac{i}{2})+\frac{(k'-1)(p+q)}{4}+\frac{3}{2}-j$
$3$-couples. It reads:
\begin{eqnarray}\label{eq.64}
{\tilde{\Lambda}'}_{even} &=&
\frac{(k'-1)(p+q+1)}{4}\biggl(\frac{(p+q-1)}{4}[2E(\frac{i}{2})+3
\notag \\ &+&\frac{(k'-1)(p+q)}{2}]\notag \\ & - &
\frac{1}{6}(p+q)(p+q-1)\biggr).
\end{eqnarray}
One deduces
$\tilde{\Lambda}'$:
\begin{eqnarray}\label{eq.65} \tilde{\Lambda}'
&=& \frac{(k'-1)(3k'-4)(p+q)^{3}}{48}\notag \\
&+&\frac{(k'-1)(p+q)^{2}}{4}[E(\frac{i}{2})+1]+\frac{(k'^{2}-1)(p+q)^{2}}{16}\notag
\\ &+&\frac{(k'-1)(p+q)}{4}[E(\frac{i}{2})+\frac{19}{12}].
\end{eqnarray}
Determination of the number of switch-off couples is
made on the same way as for cases
 $A$ and $B$. For $i$ even and $k'$ odd,
$K=(k'-1)(p+q)+i$ is even, one obtains $\Lambda_{2}$ with
(\ref{eq.48}) and (\ref{eq.49}), thus: \begin{eqnarray}\label{eq.66}
\alpha &=& \frac{(k'-1)(k'-2)(2k'-3)(p+q)^{3}}{48}\notag \\
&+&\frac{(k'-1)(p+q)^{2}}{8}[(k'-1)(3+i)-2E(\frac{i}{2})-\frac{k'}{2}\notag
\\ &-& \frac{5}{2}]+\frac{(k'-1)(p+q)}{8}[i(i+6)+\frac{11}{2}\notag
\\ &-&2E(\frac{i}{2})].
\end{eqnarray}
(\ref{eq.66}) is divisible by $p+q$, $\tilde{\Lambda}''$ (Eq.
(\ref{eq.42})) canceling $R$ (Eq.(\ref{eq.49})). The number of
$3$-couples present without multiplicity on $IS$ is given by Eq.:
\begin{eqnarray}\label{eq.52}
\tilde{\alpha}_{2} &=&
\frac{[E(\frac{i}{2})+1][E(\frac{i}{2})+2]}{2}+\frac{(k'-1)(p+q)}{2}[E(\frac{i}{2})
\notag \\ &+&\frac{(k'-1)(p+q)}{4}+\frac{3}{2}]. \end{eqnarray} One
deduces: \begin{eqnarray}\label{eq.69} \Delta_{2} &=&
\frac{[E(\frac{i}{2})+1][E(\frac{i}{2})+2]}{2} \notag \\ &+&
\frac{k'(k'-1)(p+q)}{48}[(2k'-1)(p+q)+3(2i+5)] \notag \\ \notag &+&
\frac{(k'-1)}{8}[i(i+6)+8] \notag \\ &-&
\frac{(k'-1)}{16}[4E(\frac{i}{2})-(p+q)+5]. \end{eqnarray} It is Eq.
(\ref{theor.3c}) for $i$ even.

\item If $i$ is odd, one counts $\frac{p+q+1}{2}$ classes of odd
multiplicity with populations $\tilde{\Lambda}_{2j+1}$ ($0\leq j\leq
\frac{p+q-1}{2}$), each of these classes appearing exactly one time
on each of the $\frac{(k'-1)}{2}$ odd classes of couples $C_{k'-1}$,
$...$, $C_{1}$. Moreover, for a given cumulative multiplicity
$\mu=2j+1$, there is
$E(\frac{i}{2})+\frac{(k'-1)(p+q)}{4}+\frac{3}{2}-j$ $3$-couples.
One has: \begin{eqnarray}\label{eq.70} {\tilde{\Lambda}'}_{odd} &=&
\frac{(k'-1)(p+q+1)}{24}\biggl(3[2\epsilon
E(\frac{i}{2})+3\epsilon]\notag \\ &+&\frac{3(k'-1)(p+q)}{2}
-(p+q-1)(p+q) \notag \\ &+&\frac{3\epsilon(p+q-1)}{2}
+\frac{3(p+q-1)}{2}[2E(\frac{i}{2}) \notag \\
&+&3-\epsilon+\frac{(k'-1)(p+q)}{2}]\biggr). \end{eqnarray} There is
also $\frac{p+q-1}{2}$ classes of even multiplicity with populations
$\tilde{\Lambda}_{2j}$ ($1\leq j\leq \frac{p+q-1}{2}$), each of
these classes appearing exactly one time on each of these
$\frac{(k'-1)}{2}$ odd classes of couples $C_{k'-2}$, $...$,
$C_{1}$. Furthermore, for a given cumulative multiplicity $\mu=2j$,
there is $E(\frac{i}{2})+\frac{(k'+1)(p+q)}{4}+2-j$ $3$-couples. It
gives: \begin{eqnarray}\label{eq.71} {\tilde{\Lambda}'}_{even} &=&
\frac{(k'-1)(p+q-1)}{4}\biggl(\frac{(p+q-3)}{4}[2E(\frac{i}{2})
\notag \\ &+&\frac{(k'+1)(p+q)}{2}+1-\epsilon]
+(\epsilon+1)[E(\frac{i}{2}) \notag \\ &+&\frac{(k'+1)(p+q)}{4}+1]
\notag \\ &-& \frac{1}{6}(p+q-2)(p+q-3)\biggr). \end{eqnarray} One
deduces the populations $\tilde{\Lambda}$ of all the classes $C_{j}$
($j=1,\,...,\,k'$) by summing the Eqs. (\ref{eq.43}), (\ref{eq.70})
and (\ref{eq.71}), then $\Lambda_{2}$ with (\ref{eq.48}) and
(\ref{eq.50}) and $\alpha = \Lambda_{2}-\tilde{\Lambda}$. The number
of $3$-couples present without multiplicity on $IS$ being given by
Eq. (\ref{eq.52}), it follows that: \begin{eqnarray}\label{eq.73}
\Delta_{2} &=& \frac{[E(\frac{i}{2})+1][E(\frac{i}{2})+2]}{2} \notag
\\ &+& \frac{k'(k'-1)(p+q)}{48}[(2k'-1)(p+q)+3(2i+5)] \notag \\
\notag &+& \frac{(k'-1)}{8}[i(i+6)-4 E(\frac{i}{2})+1] \notag \\ &+&
\frac{(k'-1)}{16}[4E(\frac{i}{2})-(p+q)+7]. \end{eqnarray} One
obtains $\Delta_{2}$ by (\ref{theor.3c}) for $i$ odd ($\epsilon=1$).
Theorem $6$ is demonstrated.
\end{itemize}
\begin{table}[h]
\begin{center} \begin{tabular}[h]{|c|c|c|} \hline class $C_{j}$ &
$q_{3}$ & $\mu$\\ \hline $C_{k'-1}$ & $E(\frac{i}{2})+3$ & $p+q$ \\
$C_{k'-1}$ & $E(\frac{i}{2})+4$ & $p+q-2$  \\ \vdots & \vdots &
\vdots \\ $C_{k'-1}$ & $E(\frac{i}{2})+\frac{p+q}{2}+\frac{5}{2}$ &
$1$ \\ \hline $C_{k'-2}$ &
$E(\frac{i}{2})+\frac{p+q}{2}+\frac{7}{2}$ &$p+q-1$ \\ $C_{k'-2}$ &
$E(\frac{i}{2})+\frac{p+q}{2}+\frac{9}{2}$ & $p+q-3$ \\ \vdots &
\vdots & \vdots \\ $C_{k'-2}$ & $E(\frac{i}{2})+p+q+2$ & $2$ \\
\hline \vdots & \vdots & \vdots \\ \hline $C_{2}$ &
$E(\frac{i}{2})+\frac{(k'-3)(p+q)}{2}+3$ & $p+q$ \\ $C_{2}$ &
$E(\frac{i}{2})+\frac{(k'-3)(p+q)}{2}+4$ & $p+q-2$ \\ \vdots &
\vdots & \vdots \\ $C_{2}$ &
$E(\frac{i}{2})+\frac{(k'-2)(p+q)}{2}++\frac{5}{2}$ & $1$ \\ \hline
$C_{1}$ & $E(\frac{i}{2})+\frac{(k'-2)(p+q)}{2}+\frac{7}{2}$ &
$p+q-1$ \\ $C_{1}$ &
$E(\frac{i}{2})+\frac{(k'-2)(p+q)}{2}+\frac{9}{2}$ & $p+q-3$ \\
\vdots & \vdots & \vdots  \\ $C_{1}$ &
$E(\frac{i}{2})+\frac{(k'-1)(p+q)}{2}+2$ & $2$ \\ \hline
\end{tabular} \caption{Case $C$. Table giving the different classes
of $3$-couples as a function of $q_{3}$ with the cumulative
multiplicities for $k'$ and $i$ odd.}\label{table.Appendix.10}
\end{center} \end{table}


\begin{thebibliography}{00}
\bibitem{JOY1997} M. Joyeux, On resonance-type effective vibrational
Hamiltonians for CO$_{2}$ II Results, Chem. Phys. 221 (1997)
287-301.
\bibitem{JJS1999} R. Jost, M. Joyeux, Sergei Skokov
and Joel Bowman, Vibrational analysis of HOCl up to 98 $\%$ of the
dissociation energy with a Fermi resonance Hamiltonian, J. Chem.
Phys. 111 (1999) 6807-6820.
\bibitem{JFS2002} M. Joyeux, S. C.
Farantos, R. Schinke, Highly Excited Motion in Molecules:
Saddle-Node Bifurcations and Their Fingerprints in Vibrational
Spectra, J. Phys. Chem. A 106 (2002) 5407-5421.
\bibitem{BHLV2003}
H. Broer, I. Hoveijn, G. Lunter, G. Vegter, Bifurcations in
Hamiltonian systems, Springer LNM 1806, 2003.
\bibitem{MHO2009} K. Meyer, G. R. Hall, D. Offin, Introduction to
Hamiltonian Dynamical Systems and the N-Body Problem, Applied
Mathematical Sciences, Volume 90, Springer, 2nd edition, 2009.
\bibitem{SLJ2015} G. Saget, C. Leroy, H.-R. Jauslin,
Construction and counting of the number operators of an
$n$-degree-of-freedom normalized non-resonant vibrational
Hamiltonian (2015) (submitted).
\bibitem{BIR1927} G. D. Birkhoff,
Dynamical Systems. AMS Colloq. Publ. vol 9, Providence, RI, 1927.
\bibitem{GUS1964} F.G.
Gustavson, On Constructing Formals Integrals of a Hamiltonian System
Near an Equilibrium Point, The Astronomical Journal 71 (1966)
670-686.
\bibitem{DEP1969} A. Deprit, Canonical
transformations depending on a small parameter, Cel. Mech. 1 (1969)
12-30.
\bibitem{WAY1991} R. P. Wayne, Chemistry of Atmospheres,
2nd ed, Oxford University Press, New York, 1991.
\bibitem{SBP1998}
S. Skokov, J. M. Bowman, and K. A. Peterson, An accurate ab initio
HOCl potential energy surface, vibrational and rotational
calculations, and comparison with experiment, J. Chem. Phys. 109
(1998) 2662-2671.
\bibitem{SBM1999} S. Skokov, J. M. Bowman, and V. A.
Mandelshtam, Calculation of resonance states of non-rotating HOCl
using an accurate ab initio potential, Chem. Phys. 1 (1999)
1279-1282.
\bibitem{ARN1976} V.I.
Arnold, Mathematical Methods of Classical Mechanics,
Springer-Verlag, 1976.
\bibitem{CB1997} R. H. Cushman and L. Bates,
Global aspects of classical integrable systems, Birkhäuser, 1997.
\bibitem{EFS2005} K. Efstathiou, Metamorphoses of Hamiltonian
Systems with Symmetries, LNM 1864, Springer-Verlag, 2005.
\bibitem{HIL1893} D. Hilbert, Uber die vollen InvariantenSysteme, Math. Ann. 42 (1893) 313-373.
\bibitem{GAT2000} K. Gatermann,
Computer Algebra Methods for Equivariant Dynamical Systems, volume
LNM 1728, Springer-Verlag, 2000.
\bibitem{CLO1992} D. Cox, J.
Little, and D. O'Shea, Ideals, Varieties, and Algorithms.
Undergraduate Texts in Mathematics, Springer-Verlag, 1992.
\bibitem{JOY1999} M. Joyeux, Gustavson's procedure and the dynamics
of highly excited vibrational states, J. Chem. Phys. 109 (1998)
2111-2122.
\bibitem{JS2002} M. Joyeux, D. Sugny, Canonical
perturbation theory for highly excited dynamics, Can. J. Phys. 80
(2002) 1459-1480.
\end{thebibliography}
\end{document}